\documentclass[journal,a4paper]{IEEEtran}

\usepackage[T1]{fontenc}

\usepackage{graphicx}
\usepackage[caption=false,font=footnotesize]{subfig}
\usepackage{tikz}
\usetikzlibrary{arrows.meta,calc,positioning,decorations.pathmorphing,fit}

\usepackage{amsthm,amsmath,amssymb,bm}
\usepackage{mathrsfs}
\usepackage{amsfonts}

\newtheorem{proposition}{Proposition}

\usepackage{algorithm}
\usepackage{algpseudocode}

\usepackage{xcolor}
\usepackage{booktabs}
\usepackage{verbatim}
\usepackage{placeins}

\usepackage{cite}
\usepackage[colorlinks,linkcolor=blue,urlcolor=blue,citecolor=blue]{hyperref}

\usepackage{siunitx}
\AtBeginDocument{%
  \setlength{\abovedisplayskip}{4.5pt plus 2pt minus 1pt}%
  \setlength{\belowdisplayskip}{4.5pt plus 2pt minus 1pt}%
  \setlength{\abovedisplayshortskip}{0pt plus 1pt}%
  \setlength{\belowdisplayshortskip}{3pt plus 1pt minus 1pt}%
  \setlength{\jot}{2pt}%
  \setlength{\floatsep}{7pt plus 2pt minus 2pt}%
  \setlength{\textfloatsep}{9pt plus 2pt minus 2pt}%
  \setlength{\dblfloatsep}{7pt plus 2pt minus 2pt}%
  \setlength{\dbltextfloatsep}{9pt plus 2pt minus 2pt}%
}

\newcommand{\transpose}{\mathsf{T}}
\newcommand{\hermconj}{\mathsf{H}}
\newcommand{\real}{\operatorname{Re}}

\newcommand{\diag}{\operatorname{diag}}
\newcommand{\sinc}{\operatorname{sinc}}

\begin{document}
	
\title{Mutual-Coupling-Aware Joint Precoding and Antenna Positioning for Movable-Antenna MU-MIMO Systems}
\author{Wei~Guo,~\IEEEmembership{Member,~IEEE},
Tianyi~Liao,~\IEEEmembership{Graduate~Student~Member,~IEEE},
Zixin~Wang,~\IEEEmembership{Member,~IEEE},
Shenghui~Song,~\IEEEmembership{Senior~Member,~IEEE},
Jun~Zhang,~\IEEEmembership{Fellow,~IEEE},
Khaled~B.~Letaief,~\IEEEmembership{Fellow,~IEEE}%
\thanks{%
	The authors are with the Department of Electronic and Computer Engineering, Hong Kong University of Science and Technology, Hong Kong SAR. Emails: \mbox{eeweiguo@ust.hk}, \mbox{ty.liao@connect.ust.hk}, \mbox{eewangzx@ust.hk}, \mbox{\{eeshsong, eejzhang, eekhaled\}@ust.hk}.
}%
}

\maketitle

\begin{abstract}
\textcolor{black}{Movable antennas (MAs) improve wireless communications by adapting antenna positions to the propagation environment. However, antenna movement also changes the mutual coupling (MC) among elements of the transmit and receive arrays, which affects the performance of MA-enabled systems.} In this paper, we investigate MC-aware joint precoding and antenna positioning for a downlink multiuser multiple-input multiple-output (MU-MIMO) system with MAs at both the base station and users. We establish an MC-aware system model by combining field-response propagation with a circuit-based representation of the transmit and receive arrays. Based on this model, we formulate a weighted sum-rate (WSR) maximization problem that jointly optimizes multi-stream precoders and antenna positions. The problem is challenging because the precoders and positions are coupled through the propagation channels and position-dependent MC matrices. To address this challenge, we reformulate the problem in the antenna-current domain to to efficiently handle the MC matrices and develop a current-domain weighted minimum mean square error (WMMSE) framework for alternating optimization. Furthermore, we handle the nonconvex position updates within the framework by using majorization--minimization (MM) and solve the resulting convex subproblems with a dedicated conic alternating direction method of multipliers (ADMM) algorithm. Simulation results demonstrate that the proposed design improves the WSR over both an MC-aware fixed array and a movable-array design that ignores MC during optimization.
\end{abstract}


\section{Introduction}\label{sec:intro}
Sixth-generation (6G) wireless networks are envisioned to support diverse applications, from immersive communication to intelligent connected systems, with increasingly demanding data rate and connectivity requirements~\cite{letaief2019roadmap,saad2020vision,lin2024fundamentals}. \textcolor{black}{Multiuser multiple-input multiple-output (MU-MIMO) systems use multiple antennas to serve users simultaneously, with precoding at the base station (BS) and receive processing managing interference~\cite{larsson2014massive,lu2014overview}. Their performance depends on the propagation channels, but conventional arrays have fixed element positions and cannot adapt them to changing channel conditions. Recently, reconfigurable antenna systems, including movable antennas (MAs) and fluid antenna systems (FASs), have emerged as a means of addressing this limitation by adapting antenna positions to the propagation environment~\cite{ma2023mimo,new2024mimofas,tang2026unified,wang2026rethinking}. By introducing antenna position as a design dimension, these systems offer greater flexibility in using the available spatial resources. The antenna positions can be configured alongside other communication parameters to improve system performance under different operating conditions.}

\textcolor{black}{In an MA system, antenna movement changes both the propagation channels and the electromagnetic interactions within the arrays. Specifically, movement changes the phases of multipath components and reshapes the channels~\cite{zhu2024modeling,ma2023mimo}, while also changing the element separations and hence the mutual coupling (MC) within the transmit and receive arrays~\cite{liao2026rethinking,xu2026multiusercoupling}. MA system design therefore needs to account for both effects of antenna movement. In particular, MC affects the effective channels, the transmit power associated with the antenna currents, and the receive noise covariance~\cite{ivrlac2010toward,liao2026rethinking}. Antenna positions that improve the propagation channels may also change these coupling-dependent quantities, so their overall impact on performance cannot be assessed from propagation alone. In a multiuser system, these effects influence both desired signals and interference, and must therefore be considered when adapting antenna positions and precoding to serve the users. This motivates the joint design of precoding and transmit and receive antenna positions for MA-enabled MU-MIMO systems with MC.}

\subsection{Related Works}
\textcolor{black}{Existing studies relevant to this problem have explored antenna position optimization in MA-enabled MIMO systems and the effects of MC on communication performance. The former examines how position adaptation can improve communication performance, often together with transceiver design. The latter provides models for relating electromagnetic interactions to communication performance and includes recent work on position-dependent MC in movable arrays.}

\subsubsection{Optimization of MA-enabled MIMO and MU-MIMO systems:} The field-response model provides a framework for describing how antenna movement changes a multipath channel~\cite{ma2023mimo,zhu2024modeling}. Based on this model, the authors of~\cite{ma2023mimo} investigated point-to-point MIMO capacity by jointly optimizing the transmit and receive MA positions and the transmit covariance matrix. For MIMO-FAS, New \emph{et al.}~\cite{new2024mimofas} studied joint port selection and transceiver design and analyzed the associated information-theoretic performance. In multiuser communication, antenna positioning can improve spatial separation between users as well as their individual channel strengths. Zhu \emph{et al.}~\cite{zhu2023movable} demonstrated this benefit by optimizing antenna positions and receive combining for multiuser transmission.

Joint design of beamforming and antenna positions has since been studied under different multiuser objectives and system configurations. In particular, Feng \emph{et al.}~\cite{feng2024weighted} employed weighted minimum mean square error (WMMSE) and majorization--minimization (MM) methods to optimize beamforming and MA positions for weighted sum-rate (WSR) maximization. For FAS-assisted MU-MIMO, Liao \emph{et al.}~\cite{liao2026joint} combined matrix fractional programming with parallel position updates and developed a decentralized implementation to reduce computation time. The authors of~\cite{yang2025flexible} incorporated transmit and receive antenna movement into a flexible WMMSE beamforming framework, using sparse optimization to select antenna locations. Beyond instantaneous joint design, Zheng \emph{et al.}~\cite{zheng2025twotimescale} considered a two-timescale approach that adapts positions to channel statistics and beamforming to instantaneous channels. The authors of~\cite{pi2025nearfield} addressed WSR maximization with near-field propagation, while Irshad \emph{et al.}~\cite{irshad2026wideband} evaluated MA arrays in wideband multiuser operation, including hardware impairments and their interaction with user loading. These studies demonstrate the versatility of \textcolor{black}{antenna position adaptation}, but their optimization models do not explicitly account for position-dependent MC within the arrays.

\subsubsection{Mutual coupling in antenna arrays:} MC has long been studied using circuit-based communication models, which describe antenna arrays as multiport networks and connect their electromagnetic behavior to communication performance~\cite{ivrlac2010toward}. For dense holographic arrays, Yuan \emph{et al.}~\cite{yuan2023effects} examined the effects of MC on spatial degrees of freedom and antenna efficiency, while Pizzo and Lozano~\cite{pizzo2025mutual} analyzed coupling and spatial correlation and showed how antenna patterns can be designed to improve information-theoretic performance. In an MA array, coupling is not determined solely by a fixed deployment: it varies with the \textcolor{black}{antenna positions} and therefore changes during position optimization. A minimum-spacing constraint can limit coupling and prevent physical overlap, but it does not generally eliminate coupling for \textcolor{black}{arbitrary antenna types and arrangements}. Moreover, treating coupling only as an impairment can overlook the performance attainable by accounting for it in array design.

Recent work has begun to investigate this opportunity. The authors of~\cite{zhu2026couplingenhanced} incorporated MC into MA array design and employed a hybrid convex and particle swarm optimization method for sum-rate maximization, while Xu \emph{et al.}~\cite{xu2026directivity} optimized MA positions to enhance array directivity. For planar fluid arrays, Zhang \emph{et al.}~\cite{zhang2026electromagnetic} developed a current-domain electromagnetic model incorporating coupling and power constraints. The authors of~\cite{iacovelli2026coupling} related circuit-based coupling matrices to field-theoretic kernels for movable and fluid antennas on holographic surfaces. Further studies considered coupling-aware beam synthesis through continuous position refinement~\cite{xu2026offgrid} and sparse port selection~\cite{xu2026sparseports}. These array and beam synthesis studies highlight the importance of modeling the relationship between port excitation, antenna currents, and radiation when positions or active ports change. In our previous work~\cite{liao2026rethinking}, we combined field-response and circuit-based models to study MC-aware capacity maximization in point-to-point MA-MIMO systems. That work showed how antenna movement can jointly reshape the propagation channel and coupling matrices. In a multiuser downlink, Xu \emph{et al.}~\cite{xu2026multiusercoupling} considered position-dependent coupling and jointly optimized MA positions and beamforming for sum-rate maximization using greedy search, gradient descent, and fractional programming. The point-to-point formulation in~\cite{liao2026rethinking} does not involve inter-user interference, whereas the multiuser setting in~\cite{xu2026multiusercoupling} focuses on beamforming and the \textcolor{black}{positions of the transmit antennas}.

\textcolor{black}{These studies show that antenna movement can reshape both propagation channels and array coupling. In a multiuser downlink, the same transmit and receive antenna positions therefore influence desired signals and inter-user interference, as well as coupling-dependent transmit power and receive noise. Jointly accounting for these effects changes the position-design problem beyond propagation-channel adaptation. However, joint optimization of multi-stream precoders and antenna positions at both the BS and users under position-dependent MC remains an open problem.}

\subsection{Contributions}
In this paper, we investigate MC-aware WSR maximization in a downlink MA-enabled MU-MIMO system and develop an alternating framework for joint precoding and antenna positioning. Our main contributions are as follows.
\begin{itemize}
\item We establish an MC-aware MU-MIMO model by combining field-response propagation with a circuit-based representation of the transmit and receive arrays. \textcolor{black}{The model captures the coupled effects of movement on propagation channels and array interactions, including the resulting changes in accepted transmit power and receive noise.} Based on this model, we formulate a WSR maximization problem that jointly optimizes multi-stream precoders and antenna positions under power, movement region, and minimum-spacing constraints.
\item To address the position-dependent inverse square roots of the coupling matrices in the effective channels, we develop an equivalent current-domain representation using transmit antenna currents and receive port observations. This representation avoids the matrix inverse square roots while retaining MC in the transmit power and receive noise covariance. The WSR remains non-concave and couples the precoders and positions. We therefore establish an equivalent WMMSE formulation and develop an alternating framework for their joint optimization.
\item For fixed antenna positions, the precoders can be updated using WMMSE, but the antenna position subproblems remain nonconvex. We therefore develop adaptive phase graph majorization for the position subproblems. We exploit the channel phase and pairwise MC structures to construct quadratic upper bounds for the objective and transmit power. Combining these bounds with inner approximations of the spacing constraints yields convex subproblems. We develop a conic alternating direction method of multipliers (ADMM) algorithm that exploits their shared structure and closed-form cone projections to efficiently solve the subproblems.
\end{itemize}

The remainder of this paper is organized as follows. Section~\ref{sec:model_problem} present the system model and problem formulation. Section~\ref{sec:cd_wmmse_mm} introduces the proposed optimization framework. Numerical results are provided in Section~\ref{sec:simulation}, and conclusions are drawn in Section~\ref{sec:conclusion}.

\section{System Model and Problem Formulation}\label{sec:model_problem}
\begin{figure*}[t]
\centering
\includegraphics[width=\textwidth]{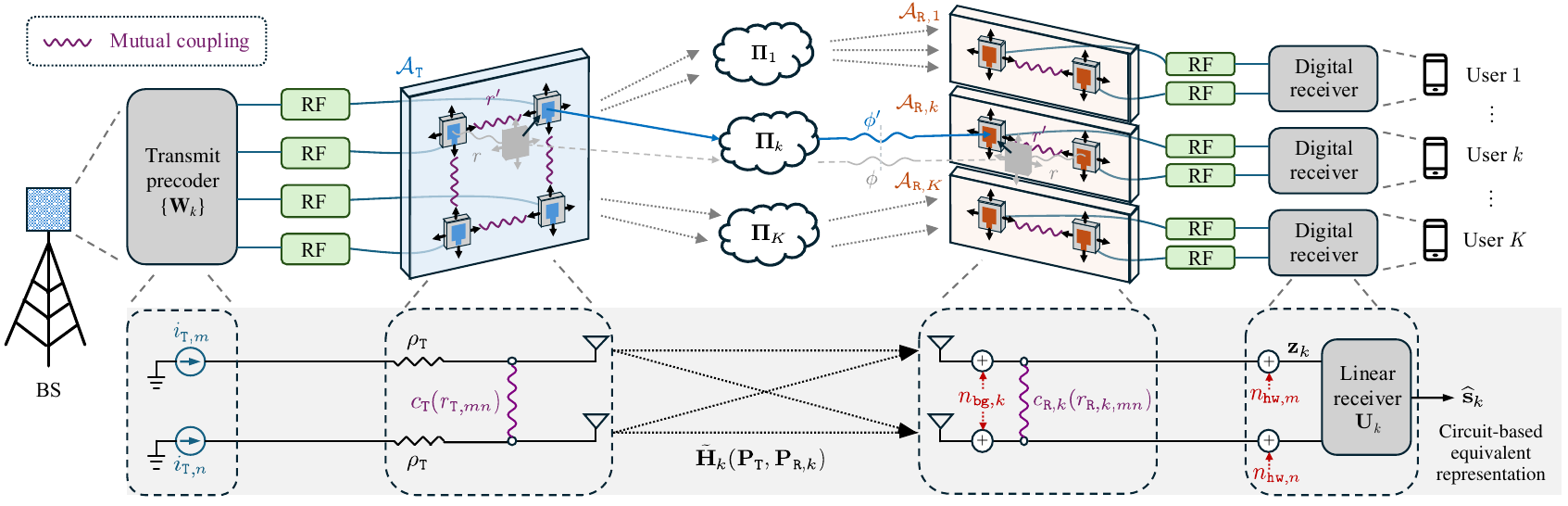}
\caption{MA-enabled MU-MIMO downlink system. In the upper planel, gray and colored traces show the highlighted configurations before and after antenna movement. The movement changes path phases ($\phi\to\phi'$) and inter-element spacings ($r\to r'$), producing channel shaping and MC variation that jointly affect the WSR. The lower panel gives the circuit-based current-domain representation using transmit antenna currents and receive port observation $\mathbf{z}_k$. The details of this representation are presented in Section~\ref{subsec:MC_Model}.}
\label{fig:system_model}
\end{figure*}

As illustrated in Fig. \ref{fig:system_model}, we consider a downlink MU-MIMO system consisting of one BS and $K$ users. The BS is equipped with $M$ movable antenna elements, while user $k$ is equipped with $N_k$ movable antenna elements. Each element has a fixed orientation and a single feed port, and we use the same index for an element and its associated port. The BS and all users are located in the far-field region of each other. The BS transmits $d_k$ independent data streams to user $k$, where $d_k\leq\min\{M,N_k\}$. The transmit antennas can move within a prescribed two-dimensional region $\mathcal{A}_{\mathtt{T}}\subset\mathbb{R}^2$. Let $\mathbf{p}_{\mathtt{T},m}\in\mathcal{A}_{\mathtt{T}}$ denote the position of transmit antenna $m$, \textcolor{black}{and collect the transmit antenna positions as}
\begin{equation}
	\mathbf{P}_{\mathtt{T}}\triangleq
	[\mathbf{p}_{\mathtt{T},1},\ldots,\mathbf{p}_{\mathtt{T},M}]
	\in\mathbb{R}^{2\times M}.
\end{equation}
Similarly, the receive antennas of user $k$ can move within a prescribed two-dimensional region $\mathcal{A}_{\mathtt{R},k}\subset\mathbb{R}^2$. With $\mathbf{p}_{\mathtt{R},k,n}\in\mathcal{A}_{\mathtt{R},k}$ denoting the position of receive antenna $n$, \textcolor{black}{the receive antenna positions of user $k$ are collected as}
\begin{equation}
	\mathbf{P}_{\mathtt{R},k}\triangleq
	[\mathbf{p}_{\mathtt{R},k,1},\ldots,\mathbf{p}_{\mathtt{R},k,N_k}]
	\in\mathbb{R}^{2\times N_k}.
\end{equation}

\subsection{Field-Response Channel Model}
\label{subsec:field_response_channel}

We adopt the field-response model to characterize the physical propagation channel~\cite{ma2023mimo,zhu2024modeling}. Let $L_{\mathtt{T},k}$ and $L_{\mathtt{R},k}$ denote the numbers of transmit and receive propagation paths associated with user $k$, respectively. The projections of the departure and arrival directions onto the local antenna movement planes are denoted by $\boldsymbol{\nu}_{\mathtt{T},k,\ell}\in\mathbb{R}^2$, $\ell=1,\ldots,L_{\mathtt{T},k}$, and $\boldsymbol{\nu}_{\mathtt{R},k,q}\in\mathbb{R}^2$, $q=1,\ldots,L_{\mathtt{R},k}$, respectively. For a projected departure direction $\boldsymbol{\nu}_{\mathtt{T}}\in\mathbb{R}^2$, the transmit steering vector is
\begin{equation}\label{eq:BS_steering_vector}
	\mathbf{a}_{\mathtt{T}}(\mathbf{P}_{\mathtt{T}},\boldsymbol{\nu}_{\mathtt{T}})
	=\left[
	e^{j\kappa\boldsymbol{\nu}_{\mathtt{T}}^{\transpose}\mathbf{p}_{\mathtt{T},1}},
	\ldots,
	e^{j\kappa\boldsymbol{\nu}_{\mathtt{T}}^{\transpose}\mathbf{p}_{\mathtt{T},M}}
	\right]^{\transpose}\in\mathbb{C}^{M},
\end{equation}
where $\kappa=2\pi/\lambda$ is the wavenumber and $\lambda$ is the wavelength. Similarly, the receive steering vector of user $k$ for a projected arrival direction $\boldsymbol{\nu}_{\mathtt{R}}\in\mathbb{R}^2$ is
\begin{equation}\label{eq:user_steering_vector}
	\mathbf{a}_{\mathtt{R},k}(\mathbf{P}_{\mathtt{R},k},\boldsymbol{\nu}_{\mathtt{R}})
	=\left[
	e^{j\kappa\boldsymbol{\nu}_{\mathtt{R}}^{\transpose}\mathbf{p}_{\mathtt{R},k,1}},
	\ldots,
	e^{j\kappa\boldsymbol{\nu}_{\mathtt{R}}^{\transpose}\mathbf{p}_{\mathtt{R},k,N_k}}
	\right]^{\transpose}\in\mathbb{C}^{N_k}.
\end{equation}

Then, the transmit field-response matrix (FRM) associated with user $k$ is constructed by stacking the steering vectors of all departure paths as
\begin{equation}\label{eq:tx_FRM}
	\begin{aligned}
	\mathbf{G}_k(\mathbf{P}_{\mathtt{T}})
	=\big[&\mathbf{a}_{\mathtt{T}}(\mathbf{P}_{\mathtt{T}},
	\boldsymbol{\nu}_{\mathtt{T},k,1}),\ldots,\\
	&\mathbf{a}_{\mathtt{T}}(\mathbf{P}_{\mathtt{T}},
	\boldsymbol{\nu}_{\mathtt{T},k,L_{\mathtt{T},k}})\big]^{\transpose}
	\in\mathbb{C}^{L_{\mathtt{T},k}\times M}.
	\end{aligned}
\end{equation}
Likewise, the receive FRM of user $k$ is given as
\begin{equation}\label{eq:rx_FRM}
	\begin{aligned}
	\mathbf{F}_k(\mathbf{P}_{\mathtt{R},k})
	=\big[&\mathbf{a}_{\mathtt{R},k}(\mathbf{P}_{\mathtt{R},k},
	\boldsymbol{\nu}_{\mathtt{R},k,1}),\ldots,\\
	&\mathbf{a}_{\mathtt{R},k}(\mathbf{P}_{\mathtt{R},k},
	\boldsymbol{\nu}_{\mathtt{R},k,L_{\mathtt{R},k}})\big]^{\transpose}
	\in\mathbb{C}^{L_{\mathtt{R},k}\times N_k}.
	\end{aligned}
\end{equation}

Let $\mathbf{\Pi}_k\in\mathbb{C}^{L_{\mathtt{R},k}\times L_{\mathtt{T},k}}$ denote the path-response matrix of user $k$, whose $(q,\ell)$-th entry characterizes the complex gain from departure path $\ell$ to arrival path $q$. The propagation channel between the BS and user $k$ is then given by
\begin{equation}\label{eq:physical_channel}
	\widetilde{\mathbf{H}}_k(\mathbf{P}_{\mathtt{T}},\mathbf{P}_{\mathtt{R},k})
	=\mathbf{F}_k(\mathbf{P}_{\mathtt{R},k})^{\hermconj}
	\mathbf{\Pi}_k
	\mathbf{G}_k(\mathbf{P}_{\mathtt{T}})
	\in\mathbb{C}^{N_k\times M}.
\end{equation}

\subsection{Circuit-Based Communication Model}\label{subsec:MC_Model}
In addition to modifying the propagation phases, antenna movement changes the MC among elements within each array. Due to these interactions, the current at one antenna port contributes to the voltages at other ports, so the antenna elements must be modeled jointly. To capture this behavior, we adopt a circuit-based model that represents each antenna array as a multiport network described by its port voltages and currents~\cite{ivrlac2010toward}. In this representation, the signal sources are connected to the transmit antenna ports through a matching network, and the source voltages produce currents at these ports. These currents generate electromagnetic waves that propagate through the environment and induce voltages at the receive antenna ports. The receive matching network then transforms these induced voltages into the output signals observed by the receivers.

Specifically, the interactions within the arrays can be characterized by the transmit MC matrix $\mathbf{C}_{\mathtt{T}}(\mathbf{P}_{\mathtt{T}})\in\mathbb{C}^{M\times M}$ and the receive MC matrix $\mathbf{C}_{\mathtt{R},k}(\mathbf{P}_{\mathtt{R},k})\in\mathbb{C}^{N_k\times N_k}$. These matrices represent the normalized radiation resistance of the arrays, with off-diagonal entries describing the mutual interactions between distinct elements. Their entries are modeled as
\begin{align}
	[\mathbf{C}_{\mathtt{T}}(\mathbf{P}_{\mathtt{T}})]_{m,n}
	&=c_{\mathtt{T}}\!\left(\|\mathbf{p}_{\mathtt{T},m}
	-\mathbf{p}_{\mathtt{T},n}\|_2\right), \label{eq:MC_entry}\\
	[\mathbf{C}_{\mathtt{R},k}(\mathbf{P}_{\mathtt{R},k})]_{m,n}
	&=c_{\mathtt{R},k}\!\left(\|\mathbf{p}_{\mathtt{R},k,m}
	-\mathbf{p}_{\mathtt{R},k,n}\|_2\right), \label{eq:MC_entry_rx}
\end{align}
where $c_{\mathtt{T}}(\cdot)$ and $c_{\mathtt{R},k}(\cdot)$ are the coupling kernels that map the distance between two elements to the corresponding coupling coefficient. Their functional forms depend on the antenna type and radiation pattern. In this paper, we consider isotropic antenna elements\footnote{Other antenna types, such as dipoles, can be described using coupling coefficients obtained from suitable antenna models or electromagnetic simulations. The position updates developed below also apply to real, twice continuously differentiable distance kernels satisfying the stated assumptions. Coupling with additional directional dependence requires corresponding changes to the derivatives and curvature bounds.}, for which the coupling kernels are given by~\cite{pizzo2025mutual}
\begin{equation}\label{eq:MC_kernel}
	c_{\mathtt{T}}(r)=c_{\mathtt{R},k}(r)=\sinc(\kappa r),
	\qquad \sinc(x)\triangleq\frac{\sin x}{x}.
\end{equation}
Given the above coupling kernel \eqref{eq:MC_kernel}, the MC matrices $\mathbf{C}_{\mathtt{T}}(\mathbf{P}_{\mathtt{T}})$ and $\mathbf{C}_{\mathtt{R},k}(\mathbf{P}_{\mathtt{R},k})$ are real symmetric and positive definite whenever the antenna positions are pairwise distinct.

The above transmit MC matrix $\mathbf{C}_{\mathtt{T}}$ describes the radiated power associated with a given antenna current vector through the normalized radiation resistance of the array. The antenna conductors also dissipate power through ohmic loss. With an identical normalized loss resistance $\rho_{\mathtt{T}}\geq0$ at each transmit element, the normalized total resistance matrix of the transmit array includes both contributions and is given by~\cite{pizzo2025mutual}
\begin{equation}\label{eq:tx_power_metric}
	\mathbf{R}_{\mathtt{T}}(\mathbf{P}_{\mathtt{T}})
	\triangleq \mathbf{C}_{\mathtt{T}}(\mathbf{P}_{\mathtt{T}})+\rho_{\mathtt{T}}\mathbf{I}_M,
	\qquad \rho_{\mathtt{T}}\geq 0.
\end{equation}
Here, $\rho_{\mathtt{T}}=0$ corresponds to the absence of ohmic loss, and $\mathbf{R}_{\mathtt{T}}\succ\mathbf{0}$ for all $\rho_{\mathtt{T}}\geq0$. We omit the \textcolor{black}{position arguments} below for notational brevity.

Let $\mathbf{s}_k\in\mathbb{C}^{d_k}$ denote the data symbol vector intended for user $k$. We assume independent Gaussian data streams, with $\mathbf{s}_k\sim\mathcal{CN}(\mathbf{0},\mathbf{I}_{d_k})$ and $\mathbb{E}[\mathbf{s}_k\mathbf{s}_j^{\hermconj}]=\mathbf{0}$ for $j\neq k$. The BS applies the precoding matrix $\mathbf{W}_k\in\mathbb{C}^{M\times d_k}$ to form the source signal as
\begin{equation}\label{eq:baseband_signal}
	\mathbf{x}=\sum_{j=1}^{K}\mathbf{W}_j\mathbf{s}_j.
\end{equation}
In the circuit model, this precoded transmit signal corresponds to the normalized baseband representation of the source voltages. The source signals are applied to the transmit array through a lossless matching network, which is assumed to maintain perfect power matching as the antenna positions change. Under the matching model and normalization adopted in~\cite{liao2026rethinking}, the normalized transmit antenna current is given by
\begin{equation}\label{eq:tx_antenna_current}
	\mathbf{i}_{\mathtt{T}}=\mathbf{R}_{\mathtt{T}}^{-1/2}\mathbf{x}.
\end{equation}
Therefore, the average power accepted by the transmit array is given as~\cite{ivrlac2010toward}
\begin{equation}\label{eq:radiated_power_equivalent}
	P_{\mathrm{acc}} =\mathbb{E}\!\left[\mathbf{i}_{\mathtt{T}}^{\hermconj}
	\mathbf{R}_{\mathtt{T}}\mathbf{i}_{\mathtt{T}}\right]
	=\mathbb{E}\!\left[\mathbf{x}^{\hermconj}\mathbf{x}\right]=\sum_{k=1}^{K}\|\mathbf{W}_k\|_{\mathrm{F}}^2.
\end{equation}

To describe propagation from the transmitter to the receivers, we adopt the unilateral approximation, which neglects the influence of the receive antenna currents on the transmitter~\cite{ivrlac2010toward}. This allows the transmit currents to be determined independently of the receive circuits. In the normalized coordinates, $\widetilde{\mathbf{H}}_k\mathbf{i}_{\mathtt{T}}$ represents the open-circuit voltages induced at the receive antennas, with constant scaling factors absorbed into~\eqref{eq:physical_channel}.

At user $k$, we assume that the receive matching network is lossless and maintain receive noise matching as the antenna positions change. Under the adopted circuit model and normalization, this network transforms the induced voltages by $\mathbf{C}_{\mathtt{R},k}^{-1/2}$~\cite{liao2026rethinking}. Thus, the received signal at user $k$ is given as
\begin{equation}\label{eq:received_signal}
	\begin{aligned}
	\mathbf{y}_k
	&=\mathbf{C}_{\mathtt{R},k}^{-1/2}
	\widetilde{\mathbf{H}}_k\mathbf{i}_{\mathtt{T}}+\mathbf{n}_k\\
	&=\mathbf{C}_{\mathtt{R},k}^{-1/2}
	\widetilde{\mathbf{H}}_k
	\mathbf{R}_{\mathtt{T}}^{-1/2}
	\mathbf{x}
	+\mathbf{n}_k,
	\end{aligned}
\end{equation}
where $\mathbf{y}_k$ is the normalized baseband representation of the output voltages after receive matching and before digital combining, and $\mathbf{n}_k$ denotes the additive noise in this observation. Based on \eqref{eq:received_signal}, we define the corresponding effective channel as
\begin{equation}\label{eq:effective_channel}
	\widehat{\mathbf{H}}_k
	\triangleq\mathbf{C}_{\mathtt{R},k}^{-1/2}
	\widetilde{\mathbf{H}}_k
	\mathbf{R}_{\mathtt{T}}^{-1/2}.
\end{equation}

To account for both background radiation and receiver hardware noise, we model the equivalent noise covariance referred to the receive ports before matching as
\begin{equation}\label{eq:receive_noise_covariance}
	\mathbf{R}_{n,k}
	=\sigma_{\mathrm{bg},k}^{2}\mathbf{C}_{\mathtt{R},k}
	+\sigma_{\mathrm{hw},k}^{2}\mathbf{I}_{N_k},
\end{equation}
where the first term represents correlated noise from an isotropic background field. The second term models receiver hardware noise referred to the same ports, which we assume to be independent across ports. The transformation associated with receive matching acts on both the signal and this equivalent noise. Accordingly, the noise in~\eqref{eq:received_signal} follows $\mathbf{n}_k\sim\mathcal{CN}(\mathbf{0},\widehat{\mathbf{R}}_{n,k})$, with
\begin{equation}\label{eq:normalized_noise_covariance}
	\widehat{\mathbf{R}}_{n,k}
	=\mathbf{C}_{\mathtt{R},k}^{-1/2}
	\mathbf{R}_{n,k}
	\mathbf{C}_{\mathtt{R},k}^{-1/2}.
\end{equation}
Then, the achievable rate of user $k$ is
\begin{equation}\label{eq:user_rate}
	R_k=\log\det\!\left(
	\mathbf{I}_{N_k}
	+\widehat{\mathbf{H}}_k\mathbf{W}_k\mathbf{W}_k^{\hermconj}
	\widehat{\mathbf{H}}_k^{\hermconj}\widehat{\mathbf{\Omega}}_k^{-1}
	\right),
\end{equation}
where $\widehat{\mathbf{\Omega}}_k$ is the covariance of interference and noise, given by
\begin{equation}\label{eq:interference_covariance}
	\widehat{\mathbf{\Omega}}_k
	=\sum_{j\neq k}\widehat{\mathbf{H}}_k
	\mathbf{W}_j\mathbf{W}_j^{\hermconj}
	\widehat{\mathbf{H}}_k^{\hermconj}
	+\widehat{\mathbf{R}}_{n,k}.
\end{equation}

\subsection{Problem Formulation}\label{subsec:problem}
\textcolor{black}{In this paper, our goal is to maximize the WSR. Since the transmit and receive antenna positions jointly determine the propagation channels and array coupling, which in turn affect the effective channels and receive noise under the accepted-power constraint. Therefore, we need to jointly optimize the multi-stream precoders and antenna positions, leading to the following problem}
\begin{subequations}\label{prob:wsr_maximization}
	\begin{align}
		\max_{\substack{\{\mathbf{W}_k\},\mathbf{P}_{\mathtt{T}},\\
		\{\mathbf{P}_{\mathtt{R},k}\}}}
		\quad & \sum_{k=1}^{K}\alpha_k R_k \label{prob:obj}\\
		\text{s.t.}\quad
		& \sum_{k=1}^{K}\|\mathbf{W}_k\|_{\mathrm{F}}^2\leq P_{\max}, \label{prob:constraint_tx_power}\\
		& \mathbf{p}_{\mathtt{T},m}\in\mathcal{A}_{\mathtt{T}},
		\quad m=1,\ldots,M, \label{prob:constraint_tx_region}\\
		& \|\mathbf{p}_{\mathtt{T},m}-\mathbf{p}_{\mathtt{T},n}\|_2
		\geq d_{\mathtt{T}}^{\min},
		\quad 1\leq m<n\leq M, \label{prob:constraint_minimum_spacing}\\
		& \mathbf{p}_{\mathtt{R},k,n}\in\mathcal{A}_{\mathtt{R},k},
		\quad n=1,\ldots,N_k,\ \forall k, \label{prob:constraint_rx_region}\\
		& \|\mathbf{p}_{\mathtt{R},k,m}-\mathbf{p}_{\mathtt{R},k,n}\|_2
		\geq d_{\mathtt{R},k}^{\min}, \nonumber\\[-0.5ex]
		& \hspace{6em}1\leq m<n\leq N_k,\quad \forall k. \label{prob:constraint_rx_spacing}
	\end{align}
\end{subequations}
Here, $\alpha_k\geq0$ in \eqref{prob:obj} denotes the priority weight of user $k$, and $P_{\max}$ in \eqref{prob:constraint_tx_power} is the maximum power budget at the BS. Constraints \eqref{prob:constraint_tx_region} and \eqref{prob:constraint_rx_region} restrict the transmit and receive antenna positions to their prescribed movement regions $\mathcal{A}_{\mathtt{T}}$ and $\mathcal{A}_{\mathtt{R},k}$, respectively. The parameters $d_{\mathtt{T}}^{\min}>0$ and $d_{\mathtt{R},k}^{\min}>0$ denote the minimum antenna spacings at the BS and user $k$, respectively. Existing studies often impose a minimum spacing of half a wavelength to mitigate MC and neglect its effects in the communication model. In this work, however, MC is explicitly included in the model and its effects are accounted for in the WSR optimization. Therefore, the minimum antenna spacings can be smaller than half a wavelength, provided that physical collisions are avoided. Constraints \eqref{prob:constraint_minimum_spacing} and \eqref{prob:constraint_rx_spacing} enforce these spacing requirements, with the minimum distances determined by the antenna dimensions and mechanical constraints.

Problem \eqref{prob:wsr_maximization} is quite challenging to solve because the precoding matrices and antenna positions are coupled in the non-concave WSR objective through both the desired signals and multiuser interference. Moreover, antenna movement changes not only the propagation channels but also the MC matrices, affecting both the effective channels and the receive noise covariances. In particular, the effective channels involve matrix inverse square roots that depend on the antenna positions, which complicates the derivatives required for position optimization. These difficulties motivate the reformulation and alternating optimization framework developed in the next section.

\section{The Proposed Current-Domain WMMSE with Majorization--Minimization Framework}
\label{sec:cd_wmmse_mm}
To address the challenges in solving the WSR maximization problem
\eqref{prob:wsr_maximization}, we develop a current-domain WMMSE framework with majorization--minimization (CD-WMMSE-MM). We first reformulate the problem in the current domain and establish an equivalent WMMSE formulation, which provides the basis for alternating optimization of the precoders and antenna positions. For fixed antenna positions, we derive the current-domain precoder update. For the transmit and receive antenna positions, we construct convex subproblems using
majorization--minimization and solve them using ADMM.

\subsection{Problem Reformulation}\label{subsec:problem_reformulation}
Directly solving \eqref{prob:wsr_maximization} is challenging because the effective channels involve matrix inverse square roots of the coupling matrices, which vary with the antenna positions. To facilitate optimization, we reformulate the problem in the current domain by introducing the current-domain precoding matrix and equivalent receive port observation as
\begin{align}
	\mathbf{J}_k
	&\triangleq\mathbf{R}_{\mathtt{T}}^{-1/2}\mathbf{W}_k
	\in\mathbb{C}^{M\times d_k}, \label{eq:current_precoder_definition}\\
	\mathbf{z}_k
	&\triangleq\mathbf{C}_{\mathtt{R},k}^{1/2}\mathbf{y}_k
	\in\mathbb{C}^{N_k}. \label{eq:port_observation_definition}
\end{align}
The matrix $\mathbf{J}_k$ maps the data streams to the normalized transmit antenna currents, while $\mathbf{z}_k$ expresses the receive observation in the coordinates before matching. Since $\mathbf{R}_{\mathtt{T}}$ and $\mathbf{C}_{\mathtt{R},k}$ are positive definite, both transformations are invertible.

Then, substituting \eqref{eq:current_precoder_definition} and \eqref{eq:port_observation_definition} into \eqref{eq:received_signal}, we obtain
\begin{equation}\label{eq:port_current_domain_signal}
	\mathbf{z}_k
	=\mathbf{H}_k\sum_{j=1}^{K}\mathbf{J}_j\mathbf{s}_j
	+\overline{\mathbf{n}}_k,
\end{equation}
where $\mathbf{H}_k\triangleq\widetilde{\mathbf{H}}_k(\mathbf{P}_{\mathtt{T}},\mathbf{P}_{\mathtt{R},k})$ and
\begin{equation}\label{eq:port_current_domain_noise}
	\overline{\mathbf{n}}_k
	\triangleq\mathbf{C}_{\mathtt{R},k}^{1/2}\mathbf{n}_k
	\sim\mathcal{CN}(\mathbf{0},\mathbf{R}_{n,k}).
\end{equation}

Thus, the achievable rate of user $k$ in the current domain can be expressed as
\begin{equation}\label{eq:port_current_domain_rate}
	R_k=\log\det\!\left(
	\mathbf{I}_{N_k}
	+\mathbf{H}_k\mathbf{J}_k\mathbf{J}_k^{\hermconj}
	\mathbf{H}_k^{\hermconj}\mathbf{\Omega}_k^{-1}
	\right),
\end{equation}
where $\mathbf{\Omega}_k$ is the interference-plus-noise covariance matrix in the current domain, given by
\begin{equation}\label{eq:port_current_interference_covariance}
	\mathbf{\Omega}_k\triangleq
	\sum_{j\neq k}\mathbf{H}_k\mathbf{J}_j\mathbf{J}_j^{\hermconj}
	\mathbf{H}_k^{\hermconj}+\mathbf{R}_{n,k}.
\end{equation}
Moreover, the accepted transmit power in \eqref{eq:radiated_power_equivalent} becomes
\begin{equation}\label{eq:port_current_domain_power}
	P_{\mathrm{acc}}
	=\sum_{k=1}^{K}
	\operatorname{Tr}\!\left(
	\mathbf{J}_k^{\hermconj}\mathbf{R}_{\mathtt{T}}\mathbf{J}_k
	\right).
\end{equation}

Although the above current-domain reformulation simplifies the treatment of MC matrices, directly optimizing the WSR remains difficult because the precoders and antenna positions are still coupled within the non-concave log-determinant rate expressions. We therefore use the standard WMMSE identity \cite{christensen2008weighted,shi2011iteratively,zhao2023rethinking} to replace these expressions with an equivalent formulation in terms of weighted MSEs. Introducing auxiliary receive matrices $\mathbf{U}_k\in\mathbb{C}^{N_k\times d_k}$ and positive definite weight matrices $\mathbf{V}_k\in\mathbb{C}^{d_k\times d_k}$, we obtain
\begin{equation}\label{eq:rate_variational_representation}
	R_k=\max_{\mathbf{U}_k,\,\mathbf{V}_k\succ\mathbf{0}}
	\left[\log\det\mathbf{V}_k-\operatorname{Tr}(\mathbf{V}_k\mathbf{E}_k)+d_k\right],
\end{equation}
where $\mathbf{E}_k$ is the MSE matrix of the estimate $\widehat{\mathbf{s}}_k=\mathbf{U}_k^{\hermconj}\mathbf{z}_k$, given by
\begin{equation}\label{eq:mse_matrix}
	\begin{aligned}
	\mathbf{E}_k={}&\mathbf{U}_k^{\hermconj}
	\left(\mathbf{H}_k\mathbf{S}\mathbf{H}_k^{\hermconj}
	+\mathbf{R}_{n,k}\right)\mathbf{U}_k
	-\mathbf{U}_k^{\hermconj}\mathbf{H}_k\mathbf{J}_k\\
	&-\mathbf{J}_k^{\hermconj}\mathbf{H}_k^{\hermconj}\mathbf{U}_k
	+\mathbf{I}_{d_k},
	\end{aligned}
\end{equation}
with the aggregate transmit current covariance defined as
\begin{equation}\label{eq:aggregate_current_covariance}
	\mathbf{S}\triangleq\sum_{j=1}^{K}\mathbf{J}_j\mathbf{J}_j^{\hermconj}.
\end{equation}

Summing \eqref{eq:rate_variational_representation} with weights $\alpha_k$, we can obtain the weighted MSE objective as
\begin{equation}\label{eq:matrix_wmmse_objective}
	\Psi\triangleq
	\sum_{k=1}^{K}\alpha_k\left[
	\operatorname{Tr}(\mathbf{V}_k\mathbf{E}_k)
	-\log\det(\mathbf{V}_k)-d_k
	\right].
\end{equation}
For any fixed precoders and antenna positions, optimizing the auxiliary variables in \eqref{eq:rate_variational_representation} yields
\begin{equation}\label{eq:wmmse_rate_identity}
	\min_{\{\mathbf{U}_k,\,\mathbf{V}_k\succ\mathbf{0}\}}\Psi
	=-\sum_{k=1}^{K}\alpha_kR_k.
\end{equation}
Therefore, problem \eqref{prob:wsr_maximization} can be equivalently reformulated as the following joint WMMSE minimization problem:
\begin{subequations}\label{prob:matrix_wmmse}
	\begin{align}
		\min_{\substack{\{\mathbf{J}_k,\mathbf{U}_k,\mathbf{V}_k\},\\
		\mathbf{P}_{\mathtt{T}},\{\mathbf{P}_{\mathtt{R},k}\}}}
		\quad&\Psi \label{prob:matrix_wmmse_obj}\\
		\mathrm{s.t.}\quad
		&P_{\mathrm{acc}}\leq P_{\max}, \label{prob:wmmse_power_constraint}\\
		&\eqref{prob:constraint_tx_region}-\eqref{prob:constraint_rx_spacing}, \label{prob:wmmse_geometry_constraints}\\
		&\mathbf{V}_k\succ\mathbf{0},\quad k=1,\ldots,K. \label{prob:wmmse_weight_constraint}
	\end{align}
\end{subequations}

In the following, we develop an alternating optimization algorithm to solve problem \eqref{prob:matrix_wmmse}. For fixed antenna positions, the receiver, auxiliary weight, and precoder blocks can be updated using standard WMMSE steps, as detailed in Section~\ref{subsec:current_precoder_update}. With these matrices fixed, we next perform one MM step for the transmit antenna positions while holding the receive positions fixed, as detailed in Section~\ref{subsec:tx_position_update}. We then perform one MM step for each user's receive antenna positions using the latest transmit positions, as detailed in Section~\ref{subsec:rx_position_update}. The receive-position subproblems are independent across users and can be solved in parallel. These position steps complete one iteration of the alternating algorithm.

\subsection{Current-Domain Precoder Update}\label{subsec:current_precoder_update}
In this step, we update $\{\mathbf{J}_k,\mathbf{U}_k,\mathbf{V}_k\}$ for fixed antenna positions $\overline{\mathbf P}_{\mathtt T}$ and $\{\overline{\mathbf P}_{\mathtt R,k}\}$. First, we minimize the WMMSE objective in \eqref{eq:matrix_wmmse_objective} with respect to the receive matrices $\mathbf{U}_k$ and MSE weights $\mathbf{V}_k$. For each user $k$, the receive matrix subproblem  is convex quadratic for fixed $\mathbf{V}_k$, while the MSE weight matrix subproblem minimizes $\operatorname{Tr}(\mathbf{V}_k\mathbf{E}_k)-\log\det\mathbf{V}_k$ over $\mathbf{V}_k\succ\mathbf{0}$ for fixed $\mathbf{U}_k$. Their first-order optimality condition yields the closed-form updates with given $\overline{\mathbf{J}}_k$\footnote{In the block updates, an overbar denotes a variable's latest available value, held fixed during the current update, or a quantity constructed at these fixed values.} as
\begin{align}
	\mathbf{U}_k^{\star}
	&=\left(\mathbf{H}_k\overline{\mathbf{S}}\mathbf{H}_k^{\hermconj}
	+\mathbf{R}_{n,k}\right)^{-1}\mathbf{H}_k\overline{\mathbf{J}}_k, \label{eq:update_receive_matrix}\\
	\mathbf{V}_k^{\star}
	&=(\mathbf{E}_k^{\star})^{-1}, \label{eq:update_weight_matrix}
\end{align}
where $\overline{\mathbf{S}}\triangleq\sum_{j=1}^{K}\overline{\mathbf{J}}_j\overline{\mathbf{J}}_j^{\hermconj}$, and $\mathbf{E}_k^{\star}$ is \eqref{eq:mse_matrix} evaluated at $\mathbf{U}_k^{\star}$ and $\{\overline{\mathbf{J}}_j\}$. With $\overline{\mathbf{U}}_k=\mathbf{U}_k^{\star}$, $\overline{\mathbf{V}}_k=\mathbf{V}_k^{\star}$, we define
\begin{align}
	\mathbf{A}&\triangleq
	\sum_{k=1}^{K}\alpha_k\mathbf{H}_k^{\hermconj}
	\overline{\mathbf{U}}_k\overline{\mathbf{V}}_k\overline{\mathbf{U}}_k^{\hermconj}\mathbf{H}_k
	\succeq\mathbf{0}, \label{eq:current_A}\\
	\mathbf{B}_k&\triangleq
	\alpha_k\mathbf{H}_k^{\hermconj}\overline{\mathbf{U}}_k\overline{\mathbf{V}}_k
	\in\mathbb{C}^{M\times d_k}. \label{eq:current_B}
\end{align}
Then, the current-domain precoder block becomes 
\begin{subequations}\label{prob:current_precoder}
	\begin{align}
		\min_{\{\mathbf{J}_k\}}\quad
		&\sum_{k=1}^{K}\left[
		\operatorname{Tr}(\mathbf{J}_k^{\hermconj}\mathbf{A}\mathbf{J}_k)
		-2\real\operatorname{Tr}(\mathbf{B}_k^{\hermconj}\mathbf{J}_k)
		\right] \label{prob:current_precoder_obj}\\
		\mathrm{s.t.}\quad
		&\sum_{k=1}^{K}\operatorname{Tr}\!\left(
		\mathbf{J}_k^{\hermconj}\mathbf{R}_{\mathtt{T}}\mathbf{J}_k
		\right)\leq P_{\max}. \label{prob:current_precoder_power}
	\end{align}
\end{subequations}

Obviously, problem \eqref{prob:current_precoder} is convex and satisfies Slater's condition for $P_{\max}>0$, we can obtain its optimal solutions by solving the KKT conditions. Specifically, we introduce the Lagrange multiplier $\lambda\geq0$ for the constraint \eqref{prob:wmmse_power_constraint}, and then obtain the stationarity condition as
\begin{equation}\label{eq:current_precoder_stationarity}
	(\mathbf{A}+\lambda\mathbf{R}_{\mathtt{T}})\mathbf{J}_k=\mathbf{B}_k.
\end{equation}
Since $\mathbf{R}_{\mathtt{T}}\succ\mathbf{0}$, premultiplying \eqref{eq:current_precoder_stationarity} by $\mathbf{R}_{\mathtt{T}}^{-1/2}$ and collecting factors gives
\begin{equation}\label{eq:scaled_precoder_stationarity}
	\left[\mathbf{R}_{\mathtt{T}}^{-1/2}\mathbf{A}\mathbf{R}_{\mathtt{T}}^{-1/2}
	+\lambda\mathbf{I}_M\right]\mathbf{R}_{\mathtt{T}}^{1/2}\mathbf{J}_k
	=\mathbf{R}_{\mathtt{T}}^{-1/2}\mathbf{B}_k.
\end{equation}
Then, according to \eqref{eq:scaled_precoder_stationarity}, we can obtain the current-domain precoder update as 
\begin{equation}\label{eq:update_current_precoder}
	\mathbf{J}_k(\lambda)
	=\mathbf{R}_{\mathtt{T}}^{-1/2}
	\left[\mathbf{R}_{\mathtt{T}}^{-1/2}\mathbf{A}\mathbf{R}_{\mathtt{T}}^{-1/2}
	+\lambda\mathbf{I}_M\right]^{\dagger}\mathbf{R}_{\mathtt{T}}^{-1/2}\mathbf{B}_k,
\end{equation}
where $(\cdot)^{\dagger}$ denotes the Moore--Penrose pseudoinverse. To determine $\lambda$, let the accepted transmit power associated with this update be
\begin{equation}\label{eq:current_power_definition}
	P(\lambda)\triangleq\sum_{k=1}^{K}\operatorname{Tr}\!\left(
	\mathbf{J}_k(\lambda)^{\hermconj}\mathbf{R}_{\mathtt{T}}\mathbf{J}_k(\lambda)\right).
\end{equation}
Based on the primal feasibility and complementary slackness conditions, the optimal multiplier $\lambda^\star$ is the smallest nonnegative
value satisfying the power constraint \eqref{prob:current_precoder_power}, i.e.,
\begin{equation}\label{eq:current_multiplier}
	\lambda^{\star}=\min\{\lambda\geq0:P(\lambda)\leq P_{\max}\}.
\end{equation}
Since $P(\lambda)$ is nonincreasing in $\lambda\geq0$, we can determine $\lambda^\star$ by bisection search.

However, each bisection iteration evaluates \eqref{eq:update_current_precoder} at a different trial multiplier. A direct implementation therefore requires repeatedly factorizing the multiplier-dependent matrix in this expression. To avoid these repeated factorizations, we use the Cholesky factorization of $\mathbf R_{\mathtt T}$ as 
\begin{equation}\label{eq:RT_cholesky}
 \mathbf R_{\mathtt T}=\mathbf L_{\mathtt T}\mathbf L_{\mathtt T}^{\hermconj},
\end{equation}
and then compute the eigendecomposition as
\begin{equation}\label{eq:generalized_evd}
	\mathbf{L}_{\mathtt{T}}^{-1}\mathbf{A}
	\mathbf{L}_{\mathtt{T}}^{-\hermconj}
	=\mathbf{Q}\diag(a_1,\ldots,a_M)\mathbf{Q}^{\hermconj}.
\end{equation}
Concatenate $\mathbf{B}\triangleq[\mathbf{B}_1,\ldots,\mathbf{B}_K]\in\mathbb{C}^{M\times D}$, where $D\triangleq\sum_kd_k$, and define
\begin{equation}
	\widetilde{\mathbf{B}}
	\triangleq\mathbf{Q}^{\hermconj}\mathbf{L}_{\mathtt{T}}^{-1}\mathbf{B}.
\end{equation}
Using these factors, the concatenated current-domain precoder can be expressed equivalently for $\lambda\geq0$ as
\begin{equation}\label{eq:current_precoder_evd}
	\mathbf{J}(\lambda)
	=\mathbf{L}_{\mathtt{T}}^{-\hermconj}\mathbf{Q}
	\left[\diag(a_1,\ldots,a_M)+\lambda\mathbf{I}_M\right]^{\dagger}
	\widetilde{\mathbf{B}},
\end{equation}
and the accepted transmit power in \eqref{eq:current_power_definition} becomes
\begin{equation}\label{eq:current_power_evd}
	P(\lambda)
	=\sum_{m=1}^{M}\sum_{\ell=1}^{D}
	\frac{|[\widetilde{\mathbf{B}}]_{m,\ell}|^2}
	{(a_m+\lambda)^2}.
\end{equation}
Here, terms with $a_m=\lambda=0$ are interpreted as zero. With this eigendecomposition, each bisection trial requires only evaluating the scalar power expression \eqref{eq:current_power_evd}, avoiding the direct evaluation of the precoder update \eqref{eq:update_current_precoder} and thereby reducing the computational cost.


\subsection{Transmit Antenna Position Update}
\label{subsec:tx_position_update}
In this step, we update $\mathbf{P}_{\mathtt T}$ while holding the receive antenna positions $\{\overline{\mathbf P}_{\mathtt R,k}\}$ and the matrices $\{\overline{\mathbf J}_k,\overline{\mathbf U}_k,\overline{\mathbf V}_k\}$ fixed. Thus, $\mathbf H_k=\widetilde{\mathbf H}_k(\mathbf P_{\mathtt T},\overline{\mathbf P}_{\mathtt R,k})$, whereas the receive noise covariances remain constant. The fixed current covariance is $\overline{\mathbf S}=\sum_j\overline{\mathbf J}_j\overline{\mathbf J}_j^{\hermconj}$. For notation brevity, we define
\begin{align}
	\mathbf{A}_k^{\mathrm{g}}
	&\triangleq\alpha_k\overline{\mathbf{U}}_k\overline{\mathbf{V}}_k\overline{\mathbf{U}}_k^{\hermconj}
	\succeq\mathbf{0}, \label{eq:geometry_Ak}\\
	\mathbf{Z}_k
	&\triangleq\alpha_k\overline{\mathbf{U}}_k\overline{\mathbf{V}}_k\overline{\mathbf{J}}_k^{\hermconj}
	\in\mathbb{C}^{N_k\times M}. \label{eq:geometry_Zk}
\end{align}
Substituting \eqref{eq:mse_matrix} into \eqref{eq:matrix_wmmse_objective} and omitting terms independent of both transmit and receive antenna positions, we obtain
\begin{equation}\label{eq:geometry_objective}
	\mathcal{F}=\sum_{k=1}^{K}\big[
	\operatorname{Tr}(\mathbf{A}_k^{\mathrm{g}}\mathbf{H}_k
	\overline{\mathbf{S}}\mathbf{H}_k^{\hermconj})
	-2\real\operatorname{Tr}(\mathbf{Z}_k^{\hermconj}\mathbf{H}_k)+\operatorname{Tr}(\mathbf{A}_k^{\mathrm{g}}\mathbf{R}_{n,k})
	\big].
\end{equation}
The accepted transmit power in \eqref{eq:port_current_domain_power} is
\begin{equation}\label{eq:tx_geometry_power}
	\mathcal{P}_{\mathtt{T}}(\mathbf{P}_{\mathtt{T}})
	=\operatorname{Tr}(\mathbf{C}_{\mathtt{T}}\overline{\mathbf{S}})
	+\rho_{\mathtt{T}}\operatorname{Tr}(\overline{\mathbf{S}}).
\end{equation}
The transmit antenna position subproblem of \eqref{prob:matrix_wmmse} is therefore
\begin{equation}\label{prob:tx_position_exact}
	\min_{\mathbf P_{\mathtt T}}\ \mathcal F\quad\mathrm{s.t.}\ \mathcal P_{\mathtt T}(\mathbf P_{\mathtt T})\leq P_{\max},\eqref{prob:constraint_tx_region},\ \eqref{prob:constraint_minimum_spacing}.
\end{equation}

Problem \eqref{prob:tx_position_exact} is hard to solve because of its nonconvex objective and constraints. To address this difficulty, we combine an MM upper bound of the objective \cite{hunter2004tutorial,sun2017majorization} with convex inner approximations of these constraints. At each update, we minimize the objective upper bound over an inner feasible set containing the current \textcolor{black}{antenna positions}.

Let $\mathbf p_{\mathtt T}\triangleq\operatorname{vec}(\mathbf P_{\mathtt T})\in\mathbb R^{2M}$ stack the candidate transmit antenna positions, and let $\overline{\mathbf p}_{\mathtt T}=\operatorname{vec}(\overline{\mathbf P}_{\mathtt T})$ contain their positions before the MM step. Using $\overline{\mathbf p}_{\mathtt T}$ as the expansion point, we define the displacement $\boldsymbol\delta_{\mathtt T}\triangleq\mathbf p_{\mathtt T}-\overline{\mathbf p}_{\mathtt T}\in\mathbb R^{2M}$, so that the current positions correspond to $\boldsymbol\delta_{\mathtt T}=\mathbf0$. Since $\overline{\mathbf p}_{\mathtt T}$ is fixed during the update, optimizing $\boldsymbol\delta_{\mathtt T}$ is equivalent to optimizing $\mathbf p_{\mathtt T}$. Once the MM step yields a displacement $\boldsymbol\delta_{\mathtt T}^\star$, we recover the updated positions as $\mathbf p_{\mathtt T}^\star=\overline{\mathbf p}_{\mathtt T}+\boldsymbol\delta_{\mathtt T}^\star$.

To obtain a convex surrogate objective, we construct a quadratic upper bound in $\boldsymbol\delta_{\mathtt T}$. For a twice continuously differentiable scalar function $f$, if $\overline{\mathbf M}\succeq\mathbf0$ upper bounds its Hessian along each candidate segment from $\overline{\mathbf p}_{\mathtt T}$ to $\overline{\mathbf p}_{\mathtt T}+\boldsymbol\delta_{\mathtt T}$, Taylor's theorem gives
\begin{equation}\label{eq:quadratic_position_majorizer}
 	f(\overline{\mathbf p}_{\mathtt T}+\boldsymbol\delta_{\mathtt T})
 	\leq\widehat f(\boldsymbol\delta_{\mathtt T})
 	\triangleq f(\overline{\mathbf p}_{\mathtt T})
 	+\nabla f(\overline{\mathbf p}_{\mathtt T})^{\transpose}\boldsymbol\delta_{\mathtt T}
 	+\frac12\boldsymbol\delta_{\mathtt T}^{\transpose}\overline{\mathbf M}\boldsymbol\delta_{\mathtt T}.
\end{equation}
Following \eqref{eq:quadratic_position_majorizer}, we first compute the gradient of $\mathcal F$ at $\overline{\mathbf p}_{\mathtt T}$ and then construct a Hessian upper bound over the candidate update region.

\subsubsection{Objective Gradient and MM Upper Bound}
\label{subsec:phase_majorization}
The transmit positions affect $\mathcal F$ in \eqref{eq:geometry_objective} through the channel matrices $\{\mathbf H_k\}$. We first express the differential of $\mathcal F$ with respect to the channel matrices as
\begin{equation}
 \mathrm d\mathcal F
 =2\real\sum_{k=1}^{K}\operatorname{Tr}\!\left[
 \big(\mathbf A_k^{\mathrm g}\mathbf H_k\overline{\mathbf S}-\mathbf Z_k\big)^{\hermconj}
 \mathrm d\mathbf H_k\right].
\end{equation}
For brevity, define the coefficient of the channel variation as
\begin{equation}\label{eq:geometry_residual}
 \mathbf\Delta_k\triangleq\mathbf A_k^{\mathrm g}\mathbf H_k\overline{\mathbf S}-\mathbf Z_k.
\end{equation}
To compute the gradient of $\mathcal F$ with respect to the transmit positions, we next compute the partial derivatives of the channel matrices with respect to the antenna coordinates. Moving transmit antenna $m$ affects only the $m$-th column of $\mathbf H_k$. For its $a$-coordinate, where $a\in\{x,y\}$, this partial derivative is
\begin{equation}
 	\mathbf c_{k,m,a}
 	\triangleq
 	\frac{\partial(\mathbf H_k\mathbf e_m)}
 	{\partial[\mathbf p_{\mathtt T,m}]_a}
 	=j\kappa\overline{\mathbf F}_k^{\hermconj}\mathbf\Pi_k
 	\mathbf D_{\mathtt T,k,a}\mathbf g_{k,m},
\end{equation}
where $\mathbf e_m$ is the $m$-th basis vector, $\mathbf g_{k,m}=\mathbf G_k\mathbf e_m$, and $\mathbf D_{\mathtt T,k,a}$ is diagonal with entries $[\boldsymbol\nu_{\mathtt T,k,\ell}]_a$. Combining this channel derivative with the differential of $\mathcal F$ gives the $a$-th entry of the channel gradient block
\begin{equation}\label{eq:tx_channel_gradient_block}
 [\mathbf g_{\mathtt T,m}^{\mathrm{ch}}]_a
 =2\real\sum_k(\mathbf\Delta_k\mathbf e_m)^{\hermconj}\mathbf c_{k,m,a}.
\end{equation}
Then, evaluating the gradient blocks in \eqref{eq:tx_channel_gradient_block} at $\overline{\mathbf p}_{\mathtt T}$ and stacking them in antenna order gives the objective gradient as
\begin{equation}\label{eq:tx_objective_gradient}
 \overline{\mathbf g}_{\mathtt T}^{\mathcal F}
 \triangleq
 \left.\nabla_{\mathbf p_{\mathtt T}}\mathcal F
 \right|_{\mathbf p_{\mathtt T}=\overline{\mathbf p}_{\mathtt T}}
 =
 \left.
 \big[
 (\mathbf g_{\mathtt T,1}^{\mathrm{ch}})^{\transpose},
 \ldots,
 (\mathbf g_{\mathtt T,M}^{\mathrm{ch}})^{\transpose}
 \big]^{\transpose}
 \right|_{\mathbf p_{\mathtt T}=\overline{\mathbf p}_{\mathtt T}}.
\end{equation}

Next, we construct a Hessian upper bound. We first rewrite the channel-dependent part of \eqref{eq:geometry_objective} in terms of $\mathbf G_k$ as
\begin{equation}\label{eq:tx_channel_path_objective}
	\mathcal{F}_{\mathtt{T}}^{\mathrm{ch}}
	=\sum_{k=1}^{K}\left[
	\operatorname{Tr}(\mathbf{\Xi}_{\mathtt{T},k}
	\mathbf{G}_k\overline{\mathbf{S}}\mathbf{G}_k^{\hermconj})
	-2\real\operatorname{Tr}(\mathbf{T}_{\mathtt{T},k}^{\hermconj}
	\mathbf{G}_k)\right],
\end{equation}
where $\mathbf\Xi_{\mathtt T,k}\triangleq\mathbf\Pi_k^{\hermconj}\overline{\mathbf F}_k\mathbf A_k^{\mathrm g}\overline{\mathbf F}_k^{\hermconj}\mathbf\Pi_k$ and $\mathbf T_{\mathtt T,k}\triangleq\mathbf\Pi_k^{\hermconj}\overline{\mathbf F}_k\mathbf Z_k$. Since each entry of $\mathbf G_k$ is a complex exponential of a linear position phase, \eqref{eq:tx_channel_path_objective} has the real harmonic representation
\begin{equation}\label{eq:generic_harmonic_representation}
	\mathcal F_{\mathtt T}^{\mathrm{ch}}(\mathbf p_{\mathtt T})
	=c_0+\sum_{q\in\mathcal{H}}
	\real\!\left\{\chi_qe^{j\mathbf{a}_q^{\transpose}\mathbf p_{\mathtt T}}\right\},
\end{equation}
where $\mathcal H$ indexes the individual nonzero position-dependent terms obtained by expanding \eqref{eq:tx_channel_path_objective} and $c_0$ is the sum of constant terms independent of $\mathbf p_{\mathtt T}$. For each term $q\in\mathcal H$, $\chi_q$ and $\mathbf a_q$ denote its complex coefficient and phase vector, respectively.  For a term arising from the quadratic part, the corresponding coefficient and phase vector are
\begin{align}
	\chi_q&=[\mathbf{\Xi}_{\mathtt{T},k}]_{r,\ell}
	[\overline{\mathbf{S}}]_{m,n},\\
	\mathbf{a}_q&=\kappa\left(
	\mathbf{e}_m\otimes\boldsymbol{\nu}_{\mathtt{T},k,\ell}
	-\mathbf{e}_n\otimes\boldsymbol{\nu}_{\mathtt{T},k,r}\right),
	\label{eq:tx_quadratic_harmonic}
\end{align}
respectively, and for a term arising from the linear part, the corresponding coefficient and phase vector are
\begin{equation}\label{eq:tx_linear_harmonic}
	\chi_q=-2[\mathbf{T}_{\mathtt{T},k}]_{\ell,m}^{*},
	\qquad
	\mathbf{a}_q=\kappa\mathbf{e}_m\otimes
	\boldsymbol{\nu}_{\mathtt{T},k,\ell},
\end{equation}
respectively. For each harmonic $h_q(\mathbf p_{\mathtt T})\triangleq|\chi_q|\cos(\mathbf a_q^{\transpose}\mathbf p_{\mathtt T}+\arg\chi_q)$, the Hessian is
\begin{equation}\label{eq:harmonic_exact_hessian_appendix}
 \nabla^2h_q(\mathbf p_{\mathtt T})
 =- |\chi_q|\cos(\mathbf a_q^{\transpose}\mathbf p_{\mathtt T}+\arg\chi_q)
 \mathbf a_q\mathbf a_q^{\transpose}.
\end{equation}
The position dependence of this Hessian lies in the scalar cosine term. We bound that term over the phases allowed by the aperture constraints, assuming compact convex apertures. Partition $\mathbf a_q$ into $2\times1$ blocks $\{\mathbf a_{q,m}\}$ corresponding to the antenna positions, and define
\begin{equation}\label{eq:aperture_phase_extrema}
 \begin{aligned}
 \theta_q^-&\triangleq\arg(\chi_q)+\sum_{m=1}^{M}
 \min_{\mathbf x\in\mathcal A_{\mathtt T}}\mathbf a_{q,m}^{\transpose}\mathbf x,\\
 \theta_q^+&\triangleq\arg(\chi_q)+\sum_{m=1}^{M}
 \max_{\mathbf x\in\mathcal A_{\mathtt T}}\mathbf a_{q,m}^{\transpose}\mathbf x.
 \end{aligned}
\end{equation}
Then, the phases over the aperture are contained in
\begin{equation}\label{eq:reachable_phase_interval}
 \overline\Theta_q\triangleq[\theta_q^-,\,\theta_q^+].
\end{equation}
The maximum of $- |\chi_q|\cos\theta$ over this interval therefore bounds the scalar in \eqref{eq:harmonic_exact_hessian_appendix}. Since $\mathbf a_q\mathbf a_q^{\transpose}\succeq\mathbf0$, a nonnegative upper bound on this scalar gives a positive semidefinite upper bound on the harmonic's Hessian. The following proposition sums these bounds and gives the resulting MM upper bound.

\begin{proposition}\label{prop:phase_majorizer}
	For $\mathcal F_{\mathtt T}^{\mathrm{ch}}$ in \eqref{eq:generic_harmonic_representation}, let $\overline\mu_q\triangleq\max_{\theta\in\overline\Theta_q}\{0,-|\chi_q|\cos\theta\}$ and $\overline{\mathbf M}_{\mathrm{ph}}\triangleq\sum_{q\in\mathcal H}\overline\mu_q\mathbf a_q\mathbf a_q^{\transpose}$. Then, for every $\mathbf p_{\mathtt T}\in\mathcal A_{\mathtt T}^{M}$, the Hessian satisfies
	\begin{equation}\label{eq:phase_hessian_upper_bound}
		\nabla^2\mathcal F_{\mathtt T}^{\mathrm{ch}}(\mathbf p_{\mathtt T})
		\preceq\overline{\mathbf M}_{\mathrm{ph}}.
	\end{equation}
	Consequently, for every $\boldsymbol\delta_{\mathtt T}$ such that $\overline{\mathbf p}_{\mathtt T}+\boldsymbol\delta_{\mathtt T}\in\mathcal A_{\mathtt T}^{M}$, we have
	\begin{equation}\label{eq:phase_majorization_inequality}
		\mathcal F_{\mathtt T}^{\mathrm{ch}}(\overline{\mathbf p}_{\mathtt T}+\boldsymbol\delta_{\mathtt T})
		\leq \mathcal F_{\mathtt T}^{\mathrm{ch}}(\overline{\mathbf p}_{\mathtt T})
		+\nabla\mathcal F_{\mathtt T}^{\mathrm{ch}}(\overline{\mathbf p}_{\mathtt T})^{\transpose}
		\boldsymbol\delta_{\mathtt T}+\frac{1}{2}\boldsymbol\delta_{\mathtt T}^{\transpose}
		\overline{\mathbf{M}}_{\mathrm{ph}}\boldsymbol\delta_{\mathtt T}.
	\end{equation}
\end{proposition}
\begin{IEEEproof}
	Please refer to Appendix \ref{append:phase_majorizer_proof}.
\end{IEEEproof}

Hence, combining the bound \eqref{eq:phase_hessian_upper_bound} in Proposition \ref{prop:phase_majorizer} with the objective gradient $\overline{\mathbf g}_{\mathtt T}^{\mathcal F}$ in \eqref{eq:tx_objective_gradient}, we obtain the MM upper bound, where $\overline{\mathcal F}\triangleq\mathcal F(\overline{\mathbf p}_{\mathtt T})$, as
\begin{equation}\label{eq:tx_objective_model_bound}
 \mathcal F(\mathbf p_{\mathtt T})\leq \widehat{\mathcal F}_{\mathtt T}(\boldsymbol\delta_{\mathtt T})\triangleq \overline{\mathcal F}
 +(\overline{\mathbf g}_{\mathtt T}^{\mathcal F})^{\transpose}
  \boldsymbol\delta_{\mathtt T}+\frac12\boldsymbol\delta_{\mathtt T}^{\transpose}\overline{\mathbf M}_{\mathrm{ph}}\boldsymbol\delta_{\mathtt T},
\end{equation}

\subsubsection{Power Gradient and Convex Inner Approximation}
\label{subsec:mc_graph_majorization}
The accepted transmit power in \eqref{eq:tx_geometry_power} depends on the transmit positions through the pairwise distances in $\mathbf C_{\mathtt T}$, making the power constraint generally nonconvex. To address this, we construct a convex upper bound on the power function and impose the power limit on this bound, yielding a convex inner approximation of the constraint.

Let $r_{mn}\triangleq\|\mathbf p_{\mathtt T,m}-\mathbf p_{\mathtt T,n}\|_2$ denote the distance in \eqref{eq:MC_entry}, and let $a_{\mathtt T,mn}\triangleq\real[\overline{\mathbf S}]_{mn}$. For distinct antenna positions, the gradient of $\mathcal{P}_{\mathtt{T}}$ w.r.t. antenna $m$ is computed as
\begin{equation}\label{eq:coupling_graph_gradient}
 \nabla_{\mathbf p_{\mathtt T,m}}\mathcal P_{\mathtt T}
 =2\sum_{n\neq m}a_{\mathtt T,mn}c_{\mathtt T}'(r_{mn})\mathbf u_{mn},
\end{equation}
where $\mathbf u_{mn}\triangleq(\mathbf p_{\mathtt T,m}-\mathbf p_{\mathtt T,n})/r_{mn}$. Evaluating these gradient blocks at $\overline{\mathbf p}_{\mathtt T}$ and stacking them in antenna order gives $\overline{\mathbf g}_{\mathtt T}^{\mathcal P}$.

Next, we bound the power Hessian over the candidate update region. We replace the nonconvex minimum-spacing constraints with their tangent inner approximations. Let $\overline{\mathbf d}_{mn}\triangleq\overline{\mathbf p}_{\mathtt T,m}-\overline{\mathbf p}_{\mathtt T,n}$ and impose
\begin{equation}\label{eq:spacing_inner_approximation}
 2(\overline{\mathbf d}_{mn})^{\transpose}
 (\mathbf p_{\mathtt T,m}-\mathbf p_{\mathtt T,n})
 -\|\overline{\mathbf d}_{mn}\|_2^2
 \geq(d_{\mathtt T}^{\min})^2.
\end{equation}
This constraint implies the original spacing constraint by convexity of the squared norm. Let $\overline r_{mn}\triangleq\|\overline{\mathbf d}_{mn}\|_2$ and $d^{\max}\triangleq\operatorname{diam}(\mathcal A_{\mathtt T})$. By the Cauchy--Schwarz inequality, the tangent and aperture constraints bound each pairwise distance as
\begin{equation}\label{eq:reachable_distance_interval}
 r_{mn}\in\overline{\mathcal R}_{mn}
 \triangleq\left[
 \frac{\overline r_{mn}^{\,2}+(d_{\mathtt T}^{\min})^2}
 {2\overline r_{mn}},\ d^{\max}\right].
\end{equation}
Each pairwise Hessian contribution has radial and tangential eigenvalues $2a_{\mathtt T,mn}c_{\mathtt T}''(r_{mn})$ and $2a_{\mathtt T,mn}c_{\mathtt T}'(r_{mn})/r_{mn}$, respectively. The following proposition bounds these eigenvalues over $\overline{\mathcal R}_{mn}$ and combines the resulting pairwise bounds.

\begin{proposition}\label{prop:coupling_majorizer}
	For every edge $m<n$, define
	\begin{equation}\label{eq:adaptive_edge_curvature}
		\overline{\omega}_{mn}
		\triangleq\max_{r\in\overline{\mathcal{R}}_{mn}}
		\left\{0,\ 2a_{\mathtt T,mn}c_{\mathtt T}''(r),\
		2a_{\mathtt T,mn}\frac{c_{\mathtt T}'(r)}{r}\right\},
	\end{equation}
	and
	\begin{equation}\label{eq:adaptive_graph_majorizer}
		\overline{\mathbf{M}}_{\mathtt T}^{\mathcal P}
		\triangleq\sum_{m<n}\overline{\omega}_{mn}
		(\mathbf{e}_m-\mathbf{e}_n)
		(\mathbf{e}_m-\mathbf{e}_n)^{\transpose}
		\otimes\mathbf{I}_2.
	\end{equation}
	For every $\boldsymbol\delta_{\mathtt T}$ such that the pairwise distances along the entire segment from $\overline{\mathbf p}_{\mathtt T}$ to $\overline{\mathbf p}_{\mathtt T}+\boldsymbol\delta_{\mathtt T}$ lie in their respective intervals $\overline{\mathcal R}_{mn}$,
	\begin{equation}\label{eq:coupling_majorization_inequality}
		\mathcal P_{\mathtt T}(\overline{\mathbf p}_{\mathtt T}+\boldsymbol\delta_{\mathtt T})
		\leq \mathcal P_{\mathtt T}(\overline{\mathbf p}_{\mathtt T})
		+(\overline{\mathbf g}_{\mathtt T}^{\mathcal P})^{\transpose}
		\boldsymbol\delta_{\mathtt T}
		+\frac{1}{2}\boldsymbol\delta_{\mathtt T}^{\transpose}
		\overline{\mathbf{M}}_{\mathtt T}^{\mathcal P}\boldsymbol\delta_{\mathtt T}.
	\end{equation}
\end{proposition}
\begin{IEEEproof}
	Please refer to Appendix \ref{append:coupling_majorizer_proof}.
\end{IEEEproof}

Hence, combining Proposition~\ref{prop:coupling_majorizer} with the power gradient $\overline{\mathbf g}_{\mathtt T}^{\mathcal P}$ gives the convex power upper bound over the update domain defined below:
\begin{equation}\label{eq:tx_power_model}
 \widehat{\mathcal P}_{\mathtt T}(\boldsymbol\delta_{\mathtt T})
 \triangleq \overline{\mathcal P}_{\mathtt T}
 +(\overline{\mathbf g}_{\mathtt T}^{\mathcal P})^{\transpose}
 \boldsymbol\delta_{\mathtt T}
 +\frac12\boldsymbol\delta_{\mathtt T}^{\transpose}
 \overline{\mathbf M}_{\mathtt T}^{\mathcal P}
 \boldsymbol\delta_{\mathtt T},
\end{equation}
where $\overline{\mathcal P}_{\mathtt T}\triangleq\mathcal P_{\mathtt T}(\overline{\mathbf p}_{\mathtt T})$.  Finally, combining the objective upper bound \eqref{eq:tx_objective_model_bound}, the power upper bound \eqref{eq:tx_power_model}, and the position update domain \eqref{eq:spacing_inner_approximation}, we obtain the convex surrogate of problem \eqref{prob:tx_position_exact} as
\begin{subequations}\label{prob:tx_geometry_update}
 \begin{align}
  \min_{\boldsymbol\delta_{\mathtt T}}\quad&
  \widehat{\mathcal F}_{\mathtt T}(\boldsymbol\delta_{\mathtt T})
  \label{prob:tx_geometry_objective}\\
  \mathrm{s.t.}\quad&
  \widehat{\mathcal P}_{\mathtt T}(\boldsymbol\delta_{\mathtt T})
  \leq P_{\max}, \label{prob:tx_geometry_surrogate_power}\\
  &\overline{\mathbf p}_{\mathtt T}+\boldsymbol\delta_{\mathtt T}
  \in\overline{\mathcal D}_{\mathtt T},
  \label{prob:tx_geometry_domain}
 \end{align}
\end{subequations}
with
\begin{equation}\label{eq:iteration_geometry_domain}
 \overline{\mathcal D}_{\mathtt T}
 \triangleq\left\{\mathbf p_{\mathtt T}:
 \begin{array}{l}
 \mathbf p_{\mathtt T,m}\in\mathcal A_{\mathtt T},\quad\forall m,\\
 \eqref{eq:spacing_inner_approximation},\quad\forall m<n
 \end{array}\right\}.
\end{equation}
Now, problem \eqref{prob:tx_geometry_update} is a convex optimization problem and can be solved using standard convex optimization methods~\cite{boyd2004convex}. To efficiently solve this problem, we develop a conic ADMM method in Section~\ref{subsec:conic_admm}.

\subsection{Receive Antenna Position Update}\label{subsec:rx_position_update}
After the transmit MM step, we hold the updated transmit positions $\overline{\mathbf P}_{\mathtt T}$ and the matrices $\{\overline{\mathbf J}_k,\overline{\mathbf U}_k,\overline{\mathbf V}_k\}$ fixed and perform one MM step for each user's receive positions. For user $k$, $\mathbf H_k=\widetilde{\mathbf H}_k(\overline{\mathbf P}_{\mathtt T},\mathbf P_{\mathtt R,k})$ and $\mathbf R_{n,k}$ both vary with $\mathbf P_{\mathtt R,k}$. This position block affects only the $k$-th summand of \eqref{eq:geometry_objective}, which we denote by
\begin{equation}\label{eq:rx_position_objective}
 \begin{aligned}
 \mathcal F_{\mathtt R,k}
 ={}&\operatorname{Tr}(\mathbf A_k^{\mathrm g}\mathbf H_k\overline{\mathbf S}\mathbf H_k^{\hermconj})
 -2\real\operatorname{Tr}(\mathbf Z_k^{\hermconj}\mathbf H_k)\\
 &+\operatorname{Tr}(\mathbf A_k^{\mathrm g}\mathbf R_{n,k}).
 \end{aligned}
\end{equation}
The receive-position subproblem for user $k$ is
\begin{equation}\label{prob:rx_position_exact}
 \begin{aligned}
 \min_{\mathbf P_{\mathtt R,k}}\quad&\mathcal F_{\mathtt R,k}\\
 \mathrm{s.t.}\quad&\eqref{prob:constraint_rx_region},\ \eqref{prob:constraint_rx_spacing},
 \end{aligned}
\end{equation}
where the constraints apply to user $k$. Let $\mathbf p_{\mathtt R,k}\triangleq\operatorname{vec}(\mathbf P_{\mathtt R,k})$ and $\overline{\mathbf p}_{\mathtt R,k}=\operatorname{vec}(\overline{\mathbf P}_{\mathtt R,k})$. Define the receive displacement as $\boldsymbol\delta_{\mathtt R,k}\triangleq\mathbf p_{\mathtt R,k}-\overline{\mathbf p}_{\mathtt R,k}\in\mathbb R^{2N_k}$. The accepted transmit power is independent of the receive positions, and the $K$ subproblems are mutually independent for fixed transmit positions.

The objective and minimum-spacing constraints make \eqref{prob:rx_position_exact} nonconvex. As in the transmit update, we construct an MM upper bound on the objective and a convex inner approximation of the spacing constraints. Write $\mathcal F_{\mathtt R,k}=\mathcal F_{\mathtt R,k}^{\mathrm{ch}}+\mathcal F_{\mathtt R,k}^{\mathrm{noise}}$, where the channel and noise terms consist of the first two terms and the last term of \eqref{eq:rx_position_objective}, respectively.

All channel quantities below use the updated transmit positions $\overline{\mathbf P}_{\mathtt T}$. In particular, we recompute $\mathbf\Delta_k$ in \eqref{eq:geometry_residual} using the channel after the transmit-position update. For the channel-gradient calculation, let $\mathbf D_{\mathtt R,k,a}$ be diagonal with entries $[\boldsymbol\nu_{\mathtt R,k,\ell}]_a$ and let $\mathbf f_{k,n}=\mathbf F_k\mathbf e_n$. The derivative of the $n$-th row of $\mathbf H_k$ with respect to the $a$-coordinate of receive antenna $n$ is
\begin{equation}
 \mathbf r_{k,n,a}^{\hermconj}\triangleq-j\kappa\mathbf f_{k,n}^{\hermconj}
 \mathbf D_{\mathtt R,k,a}\mathbf\Pi_k\overline{\mathbf G}_k.
\end{equation}
The corresponding channel-gradient block is
\begin{equation}\label{eq:rx_channel_gradient_block}
 [\mathbf g_{\mathtt R,k,n}^{\mathrm{ch}}]_a
 =2\real\big[\mathbf e_n^{\transpose}\mathbf\Delta_k\mathbf r_{k,n,a}\big].
\end{equation}
Substituting \eqref{eq:receive_noise_covariance} into the noise term of \eqref{eq:rx_position_objective}, we apply the same differentiation as in \eqref{eq:coupling_graph_gradient} to obtain its receive-position gradient blocks, using $\mathbf p_{\mathtt R,k,m}$, the kernel $c_{\mathtt R,k}$, and the coefficients
\begin{equation}\label{eq:rx_noise_edge_coefficients}
 a_{\mathtt R,k,mn}
 =\sigma_{\mathrm{bg},k}^{2}\real[\mathbf A_k^{\mathrm g}]_{mn}.
\end{equation}
The hardware-noise term is position independent and contributes zero gradient. Evaluating the channel and noise gradients at $\overline{\mathbf p}_{\mathtt R,k}$ and stacking their antenna blocks gives $\overline{\mathbf g}_{\mathtt R,k}^{\mathrm{ch}}$ and $\overline{\mathbf g}_{\mathtt R,k}^{\mathrm{noise}}$, respectively. Adding these two contributions gives the objective gradient
\begin{equation}\label{eq:rx_total_gradient}
 \overline{\mathbf g}_{\mathtt R,k}^{\mathcal F}
 =\overline{\mathbf g}_{\mathtt R,k}^{\mathrm{ch}}
 +\overline{\mathbf g}_{\mathtt R,k}^{\mathrm{noise}}.
\end{equation}

We next construct Hessian upper bounds over the convex receive update domain $\overline{\mathcal D}_{\mathtt R,k}$, obtained from \eqref{eq:iteration_geometry_domain} using the receive positions, aperture $\mathcal A_{\mathtt R,k}$, and minimum spacing $d_{\mathtt R,k}^{\min}$.

To bound the channel Hessian, we rewrite the channel-dependent objective using the fixed transmit FRM $\overline{\mathbf G}_k$. Let $\mathbf B_{\mathtt R,k}=\mathbf\Pi_k\overline{\mathbf G}_k$ and define
\begin{align}
	\mathbf{\Xi}_{\mathtt{R},k}
	&\triangleq\mathbf{B}_{\mathtt{R},k}\overline{\mathbf{S}}
	\mathbf{B}_{\mathtt{R},k}^{\hermconj},\\
	\mathbf{T}_{\mathtt{R},k}
	&\triangleq\mathbf{B}_{\mathtt{R},k}\mathbf{Z}_k^{\hermconj}.
\end{align}
The channel-dependent part of \eqref{eq:rx_position_objective} can then be written as
\begin{equation}\label{eq:rx_channel_path_objective}
	\mathcal{F}_{\mathtt{R},k}^{\mathrm{ch}}
	=\operatorname{Tr}(\mathbf{\Xi}_{\mathtt{R},k}
	\mathbf{F}_k\mathbf{A}_k^{\mathrm{g}}\mathbf{F}_k^{\hermconj})
	-2\real\operatorname{Tr}(\mathbf{T}_{\mathtt{R},k}^{\hermconj}
	\mathbf{F}_k).
\end{equation}
The channel term in \eqref{eq:rx_channel_path_objective} has the same harmonic structure as \eqref{eq:tx_channel_path_objective}. Its coefficients and phase vectors follow from the transmit expressions by replacing $(\mathbf\Xi_{\mathtt T,k},\overline{\mathbf S},\mathbf T_{\mathtt T,k},\boldsymbol\nu_{\mathtt T,k,\ell})$ with $(\mathbf\Xi_{\mathtt R,k},\mathbf A_k^{\mathrm g},\mathbf T_{\mathtt R,k},\boldsymbol\nu_{\mathtt R,k,\ell})$. Following Proposition~\ref{prop:phase_majorizer}, we obtain the channel Hessian upper bound $\overline{\mathbf M}_{\mathtt R,k}^{\mathrm{ch}}$ over the receive aperture. For the noise term, the pairwise construction in Proposition~\ref{prop:coupling_majorizer}, using $c_{\mathtt R,k}$, the coefficients in \eqref{eq:rx_noise_edge_coefficients}, and the corresponding receive-distance intervals from \eqref{eq:reachable_distance_interval}, gives $\overline{\mathbf M}_{\mathtt R,k}^{\mathrm{noise}}$.

The Hessian of $\mathcal F_{\mathtt R,k}$ is the sum of the channel and noise Hessians, so their upper bounds add. We therefore set
\begin{equation}\label{eq:rx_total_majorizer}
	\overline{\mathbf{M}}_{\mathtt{R},k}^{\mathcal{F}}
	=\overline{\mathbf{M}}_{\mathtt{R},k}^{\mathrm{ch}}
	+\overline{\mathbf{M}}_{\mathtt{R},k}^{\mathrm{noise}}.
\end{equation}
Let $\overline{\mathcal F}_{\mathtt R,k}=\mathcal F_{\mathtt R,k}(\overline{\mathbf P}_{\mathtt R,k})$. Combining the channel and noise bounds gives the MM upper bound of the entire receive objective:
\begin{equation}\label{eq:rx_objective_model}
 \begin{aligned}
 \mathcal F_{\mathtt R,k}(\mathbf P_{\mathtt R,k})
 &\leq\widehat{\mathcal F}_{\mathtt R,k}(\boldsymbol\delta_{\mathtt R,k})\\
 &\triangleq\overline{\mathcal F}_{\mathtt R,k}
 +(\overline{\mathbf g}_{\mathtt R,k}^{\mathcal F})^{\transpose}
 \boldsymbol\delta_{\mathtt R,k}\\
 &\quad+\frac12\boldsymbol\delta_{\mathtt R,k}^{\transpose}
 \overline{\mathbf M}_{\mathtt R,k}^{\mathcal F}
 \boldsymbol\delta_{\mathtt R,k}.
 \end{aligned}
\end{equation}
Minimizing this upper bound over the receive update domain gives
\begin{equation}\label{prob:rx_geometry_update}
 \begin{aligned}
 \min_{\boldsymbol\delta_{\mathtt R,k}}\quad&
 \widehat{\mathcal F}_{\mathtt R,k}(\boldsymbol\delta_{\mathtt R,k})\\
 \mathrm{s.t.}\quad&
 \overline{\mathbf p}_{\mathtt R,k}+\boldsymbol\delta_{\mathtt R,k}
 \in\overline{\mathcal D}_{\mathtt R,k}.
 \end{aligned}
\end{equation}
The $K$ receive-position subproblems are convex and can be solved in parallel.

\subsection{ADMM-Based Solution to Position Subproblems}
\label{subsec:conic_admm}
The transmit and receive position subproblems share a common conic structure. We exploit this structure to solve them using ADMM. We present the formulation and updates for the transmit displacement $\boldsymbol\delta_{\mathtt T}$, and the receive counterpart follows from the corresponding coefficients and constraints. To express the transmit subproblem in conic form, we represent the spacing and aperture constraints using affine inequalities and, for disk apertures, second-order cone (SOC) constraints. The SOC is defined as $\mathcal Q_q=\{(t,\mathbf x):\|\mathbf x\|_2\leq t\}$. The remaining transmit power constraint admits a rotated SOC representation. In particular, factoring
$\overline{\mathbf M}_{\mathtt T}^{\mathcal P}
=\mathbf L_{\mathtt P}^{\transpose}\mathbf L_{\mathtt P}$ gives
\begin{equation}\label{eq:power_rotated_soc}
 \big(P_{\max}-\overline{\mathcal P}_{\mathtt T}
 -(\overline{\mathbf g}_{\mathtt T}^{\mathcal P})^{\transpose}\boldsymbol\delta_{\mathtt T},
 \ 1,\ \mathbf L_{\mathtt P}\boldsymbol\delta_{\mathtt T}\big)\in\mathcal Q_{\mathrm r},
\end{equation}
where $\mathcal Q_{\mathrm r}=\{(u,v,\mathbf w):2uv\geq\|\mathbf w\|_2^2,\ u,v\geq0\}$. We collect these constraints into
\begin{equation}\label{prob:generic_conic_geometry}
 \min_{\boldsymbol\delta_{\mathtt T}}\quad
 \tfrac12\boldsymbol\delta_{\mathtt T}^{\transpose}\mathbf Q\boldsymbol\delta_{\mathtt T}
 +\mathbf g^{\transpose}\boldsymbol\delta_{\mathtt T}
 \quad\mathrm{s.t.}\quad\mathbf K\boldsymbol\delta_{\mathtt T}+\mathbf c\in\mathcal K,
\end{equation}
where $\mathbf Q\succeq\mathbf0$ and $\mathcal K$ is the product of the corresponding cones. For the receive subproblem, the cone product contains only the aperture and spacing blocks.

To solve \eqref{prob:generic_conic_geometry}, we introduce $\mathbf z=\mathbf K\boldsymbol\delta_{\mathtt T}+\mathbf c$ and apply two-block ADMM with a scaled dual variable $\mathbf y$ \cite{boyd2011distributed}. Let $t$ index the ADMM iterations for the fixed surrogate subproblem at outer iteration $i$. With penalty $\varrho_{\mathrm{AD}}^{(t)}>0$, define
\begin{equation}\label{eq:admm_normal_matrix}
 \mathbf R_{\mathrm{AD}}^{(t)}
 =\mathbf Q+\varrho_{\mathrm{AD}}^{(t)}\mathbf K^{\transpose}\mathbf K.
\end{equation}
A nonzero null direction of the aperture-constraint blocks would leave all aperture constraints unchanged, contradicting boundedness of the apertures. Hence $\mathbf K$ has full column rank and $\mathbf R_{\mathrm{AD}}^{(t)}\succ\mathbf0$. We obtain the updates
\begin{align}
	\boldsymbol\delta_{\mathtt T}^{(t+1)}
	={}&\left(\mathbf{R}_{\mathrm{AD}}^{(t)}\right)^{-1}
	\left[-\mathbf{g}
	+
	\varrho_{\mathrm{AD}}^{(t)}\mathbf{K}^{\transpose}
	(\mathbf{z}^{(t)}-\mathbf{c}-\mathbf{y}^{(t)})\right],
	\label{eq:admm_delta_update}\\
	\mathbf{z}^{(t+1)}
	={}&\Pi_{\mathcal{K}}\!\left(
	\mathbf{K}\boldsymbol\delta_{\mathtt T}^{(t+1)}
	+\mathbf{c}+\mathbf{y}^{(t)}\right), \label{eq:admm_cone_update}\\
	\mathbf{y}^{(t+1)}
	={}&\mathbf{y}^{(t)}
	+\mathbf{K}\boldsymbol\delta_{\mathtt T}^{(t+1)}
	+\mathbf{c}-\mathbf{z}^{(t+1)}. \label{eq:admm_dual_update}
\end{align}
The product structure of $\mathcal K$ allows the projection $\Pi_{\mathcal K}$ to be evaluated separately for each cone, with closed-form projections onto the orthant, SOC, and rotated SOC. The resulting displacement is then used to update the transmit antenna positions, and the receive positions are updated in the same manner.

Overall, the alternating optimization procedure, i.e., CD-WMMSE-MM method, is summarized in Algorithm~\ref{algo:cd_wmmse_mm}.
\begin{algorithm}
\caption{The proposed CD-WMMSE-MM method}
\label{algo:cd_wmmse_mm}
\begin{algorithmic}[1]
\Require Feasible $\mathbf P_{\mathtt T}^{(0)}$, $\{\mathbf P_{\mathtt R,k}^{(0)}\}$, and $\{\mathbf W_k^{(0)}\}$
\State Initialize $\{\mathbf J_k^{(0)}\}$ using \eqref{eq:current_precoder_definition}; set $i\gets0$
\Repeat
 \State Update $\{\mathbf U_k,\mathbf V_k\}$ using \eqref{eq:update_receive_matrix} and \eqref{eq:update_weight_matrix}
 \State Update $\{\mathbf J_k\}$ using \eqref{eq:current_precoder_evd} with scalar bisection on $\lambda$
 \State Refresh $\{\mathbf U_k,\mathbf V_k\}$ using the updated $\{\mathbf J_k\}$
 \State Construct and solve the transmit-position subproblem \eqref{prob:tx_geometry_update} by ADMM to update $\mathbf P_{\mathtt T}$
 \State Using the updated transmit positions and channels, construct and solve the receive-position subproblems \eqref{prob:rx_geometry_update} by ADMM in parallel to update $\{\mathbf P_{\mathtt R,k}\}$
 \State Set $i\gets i+1$
\Until{The relative WSR improvement satisfies the prescribed tolerance}
\State Recover $\{\mathbf W_k\}$ by $\mathbf W_k=\mathbf R_{\mathtt T}(\mathbf P_{\mathtt T})^{1/2}\mathbf J_k$
\Ensure $\{\mathbf W_k\}$, $\mathbf P_{\mathtt T}$, $\{\mathbf P_{\mathtt R,k}\}$
\end{algorithmic}
\end{algorithm}

\subsection{Convergence and Complexity Analysis}
With a feasible initialization and exact subproblem solutions, the WMMSE updates and the MM-based transmit and receive position updates preserve feasibility and do not increase the WMMSE objective. By the equivalence between the WMMSE and WSR formulations, the WSR is non-decreasing over the outer iterations. Since the WSR is bounded above, the objective value of Algorithm~\ref{algo:cd_wmmse_mm} converges. For the precoder update, the eigendecomposition-based computation costs $\mathcal O(M^3+M^2D+N_{\mathrm{bis}}MD)$, where $D=\sum_k d_k$ and $N_{\mathrm{bis}}$ is the number of bisection trials. For a fixed ADMM penalty, the transmit and receive position solves have an aggregate cost of $\mathcal O(M^3+N_{\mathrm{AD},\mathtt T}M^2+\sum_k[N_k^3+N_{\mathrm{AD},\mathtt R,k}N_k^2])$, where $N_{\mathrm{AD},\mathtt T}$ and $N_{\mathrm{AD},\mathtt R,k}$ are their iteration counts.

\section{Numerical Results}\label{sec:simulation}
In this section, we evaluate the WSR performance of the MC-aware joint precoding
and antenna positioning design in MA-enabled MU-MIMO systems.

\subsection{Simulation Setup}
We consider a \SI{28}{GHz} narrowband downlink, for which the wavelength is
approximately \SI{10.714}{mm}. The default system consists of $K=4$ users,
$M=8$ transmit antennas, and $N_k=2$ receive antennas per user. Two data streams are
sent to every user, i.e., $d_k=2$, and all user weights are set to one. The
transmit antennas move in a $1.25\lambda\times0.75\lambda$ rectangular aperture,
whereas the receive antennas of every user move in a
$0.5\lambda\times0.5\lambda$ rectangular aperture. The minimum spacing at
the BS and users is $0.2\lambda$.

The physical channel follows the clustered Rician structure used in~\cite{liao2026rethinking}. In addition to the LoS
component, three scattering clusters with eight subpaths per cluster are
generated, giving 25 paired departure/arrival paths for each user. The Rician
factor is \SI{0}{dB}, the log-normal shadowing standard deviation is
\SI{4}{dB}, and user distances are drawn between \SI{100}{m} and
\SI{300}{m}. The nominal departure angles of the four users are
$[-45^\circ,-10^\circ,25^\circ,60^\circ]$. Both the transmit and receive MC
matrices use the isotropic-element kernel
$c(r)=\sinc(2\pi r/\lambda)$~\cite{pizzo2025mutual}. The normalized transmit ohmic-loss coefficient
is $\rho_{\mathtt T}=1/73.1\simeq0.01368$. The total isolated-port noise power
is \SI{-80}{dBm}; a \SI{5}{dB} noise figure is used to divide it into the
correlated background-noise and independent receiver-hardware-noise
components in \eqref{eq:receive_noise_covariance}.

The reference SNR is defined at a distance of \SI{200}{m}, with zero
shadowing and before array and beamforming gains. For a requested reference
SNR $\gamma_{\rm ref}$, the accepted-power budget is selected as
\begin{equation}
 P_{\max}[\mathrm{dBm}]
 =\gamma_{\rm ref}[\mathrm{dB}]
 +\mathrm{PL}(\SI{200}{m})[\mathrm{dB}]
 +\sigma_{\mathrm{tot}}^2[\mathrm{dBm}],
\end{equation}
where $\mathrm{PL}(\SI{200}{m})$ is the path loss at the reference distance.
Thus, the instantaneous received SNRs still vary with user distance,
shadowing, and small-scale fading.

We compare the proposed MC-aware joint precoding and antenna positioning design with the following baselines.
\begin{itemize}
\item \textbf{MC-aware fixed array:} The transmit and receive antennas are fixed on uniform grids. We consider both conventional half-wavelength-spaced arrays and compact arrays with grid spacings chosen to satisfy the aperture and minimum-spacing constraints. For each fixed layout, we optimize the precoders and receive matrices using WMMSE~\cite{christensen2008weighted,shi2011iteratively}, with MC explicitly accounted for.
\item \textbf{MC-ignored movable array:} The transmit and receive antenna positions
are jointly optimized without accounting for MC. The resulting positions and
current-domain precoders are evaluated under the MC-aware model.
\end{itemize}

The default maximum number of outer iterations is 200. The relative WSR stopping tolerance is
$2\times10^{-4}$. The absolute and relative stopping tolerances of the inner
ADMM solver are $10^{-7}$ and $10^{-6}$, respectively, and at most
\num{30000} ADMM iterations are allowed per \textcolor{black}{position update}. Unless otherwise stated, each operating point is averaged over \num{1000}
independent channel realizations.

\subsection{Convergence Behavior}
\begin{figure}[t]
 \centering
 \includegraphics[width=0.96\columnwidth]{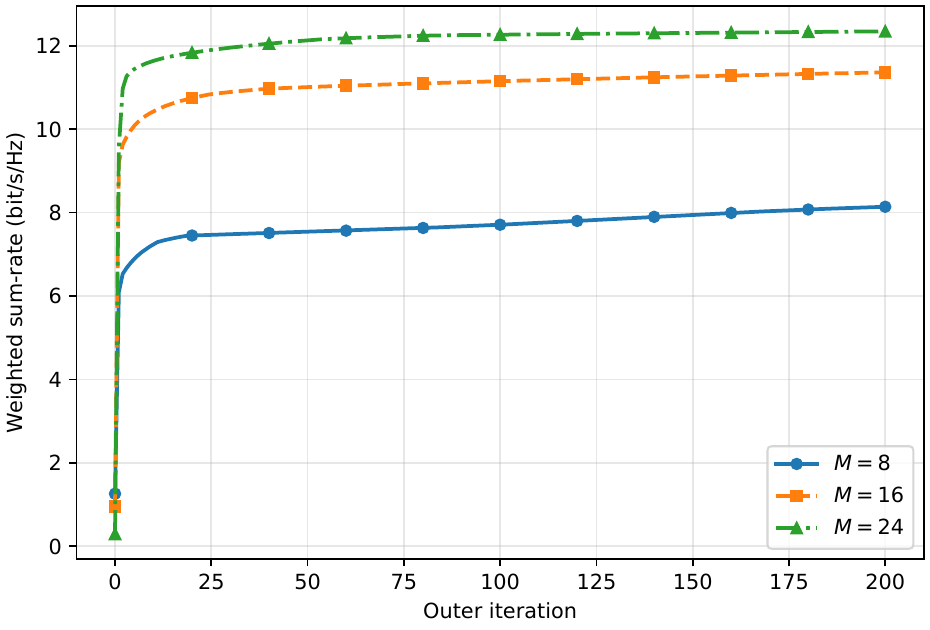}
 \caption{Convergence of CD-WMMSE-MM for different numbers of transmit antennas
 under a representative channel realization at \SI{0}{dB} reference SNR.}
 \label{fig:convergence}
\end{figure}

Fig.~\ref{fig:convergence} illustrates the WSR trajectories of CD-WMMSE-MM for one
representative channel realization at \SI{0}{dB} and
$M\in\{8,16,24\}$. All curves are recorded for 200 outer iterations without
early stopping. They exhibit a sharp increase during the first few iterations,
followed by progressively smaller improvements as the \textcolor{black}{transceiver variables and antenna positions} are refined. The WSR is nondecreasing for every array size, and the larger
arrays attain higher WSR values in this realization.

\subsection{WSR Versus Reference SNR}
\begin{figure}[t]
 \centering
 \includegraphics[width=0.96\columnwidth]{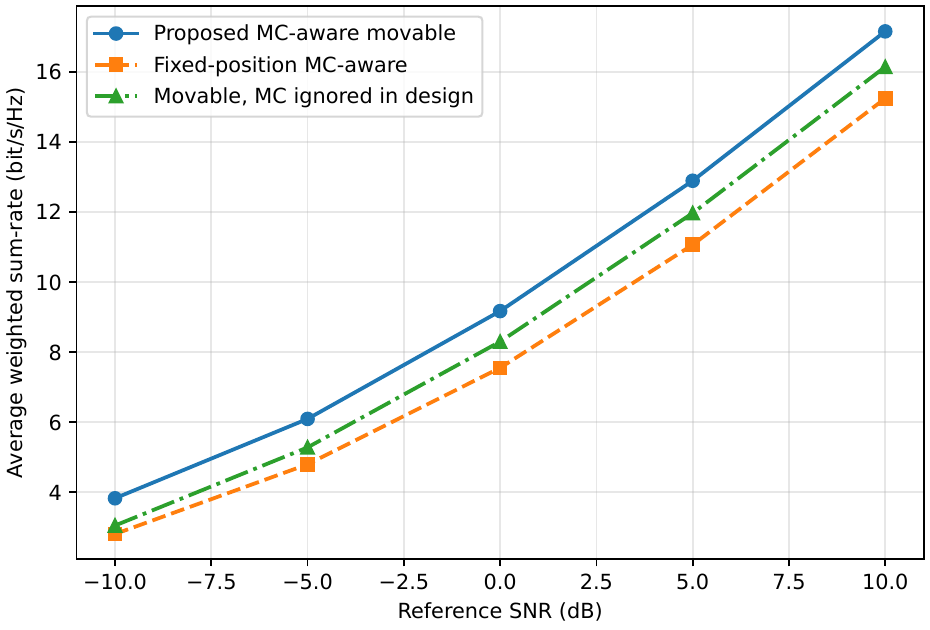}
 \caption{Average WSR versus reference SNR.}
 \label{fig:wsr_snr}
\end{figure}

Fig.~\ref{fig:wsr_snr} shows the average WSR versus the reference SNR. The MC-aware joint design consistently outperforms both baselines over the entire range
from \SI{-10}{dB} to \SI{10}{dB}. At \SI{0}{dB}, it achieves
$9.173$ bit/s/Hz, compared with $7.552$ bit/s/Hz for the conventional
half-wavelength fixed array and $8.302$ bit/s/Hz for the movable design that
ignores MC. These correspond to relative gains of $21.5\%$ and $10.5\%$,
respectively. The absolute gain over the fixed array increases from
$1.021$ bit/s/Hz at \SI{-10}{dB} to $1.913$ bit/s/Hz at \SI{10}{dB}.
Meanwhile, the persistent gap to the MC-ignored design demonstrates that
position optimization based only on the field-response channel cannot fully
exploit the movable apertures when closely spaced antenna elements are coupled.

\begin{figure*}[!t]
 \centering
 \subfloat[WSR performance w.r.t. different transmit aperture scale.\label{fig:aperture_tx}]{%
  \includegraphics[width=0.46\textwidth]{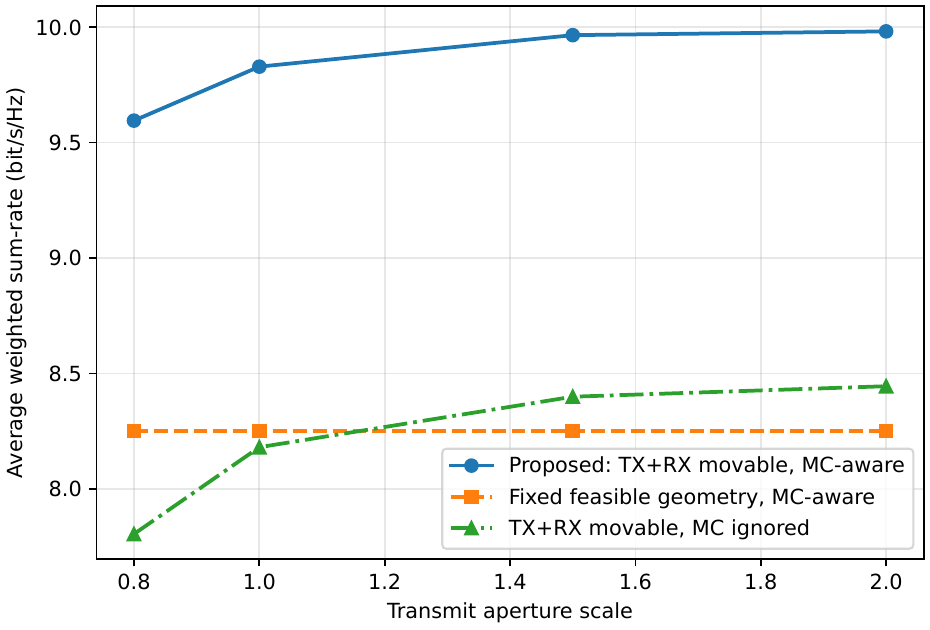}}
 \hfill
 \subfloat[WSR performance w.r.t. different receive aperture scale.\label{fig:aperture_rx}]{%
  \includegraphics[width=0.46\textwidth]{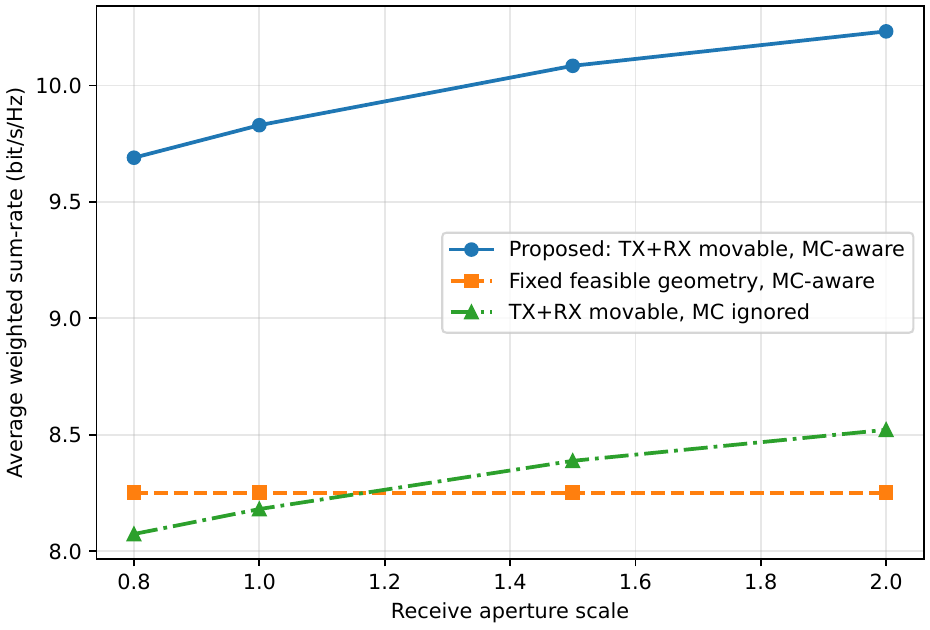}}
 \caption{Average WSR versus transmit- and receive-aperture scales at \SI{0}{dB}
 reference SNR.}
 \label{fig:aperture}
\end{figure*}

\begin{figure*}[!t]
 \centering
 \subfloat[Fixed transmit aperture.\label{fig:tx_scaling_fixed_aperture}]{%
  \includegraphics[width=0.46\textwidth]{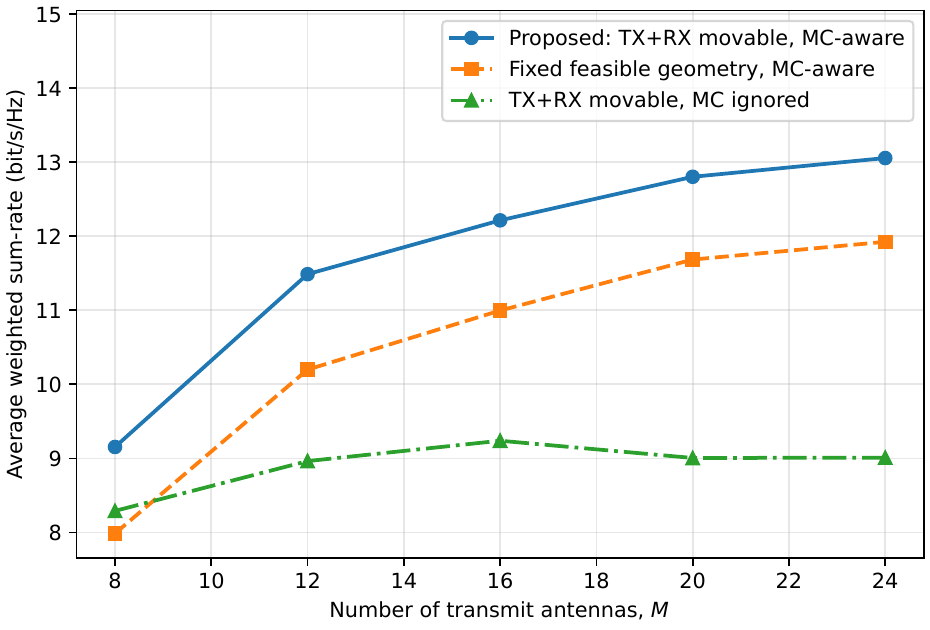}}
 \hfill
 \subfloat[Fixed transmit-antenna density.\label{fig:tx_scaling_fixed_density}]{%
  \includegraphics[width=0.46\textwidth]{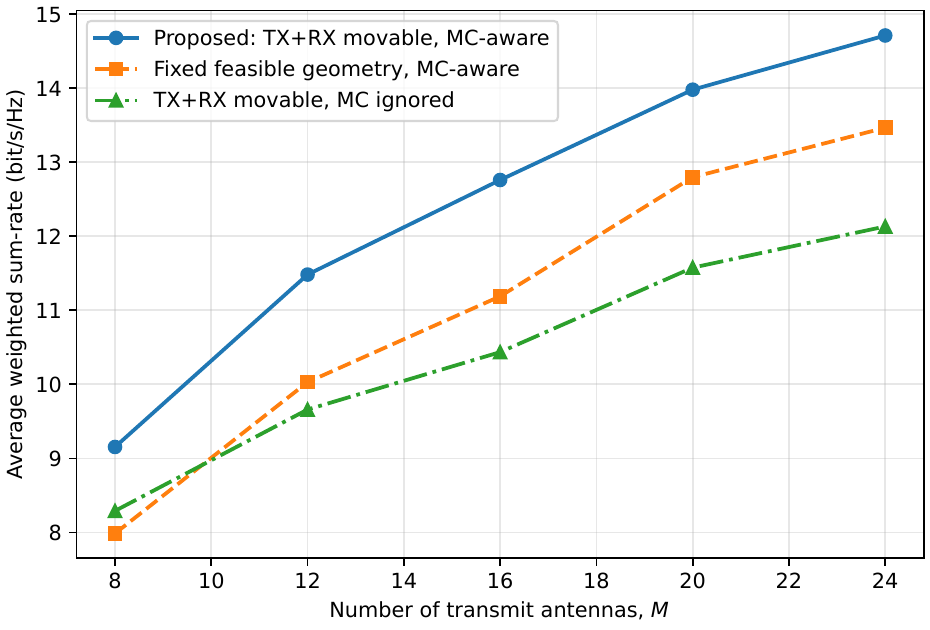}}
 \caption{Average WSR versus the number of transmit antennas at \SI{0}{dB}
 reference SNR. At fixed antenna density, both transmit-aperture dimensions
 scale as $\sqrt{M/8}$.}
 \label{fig:tx_scaling}
\end{figure*}

\begin{figure}[!t]
 \centering
 \includegraphics[width=0.96\columnwidth]{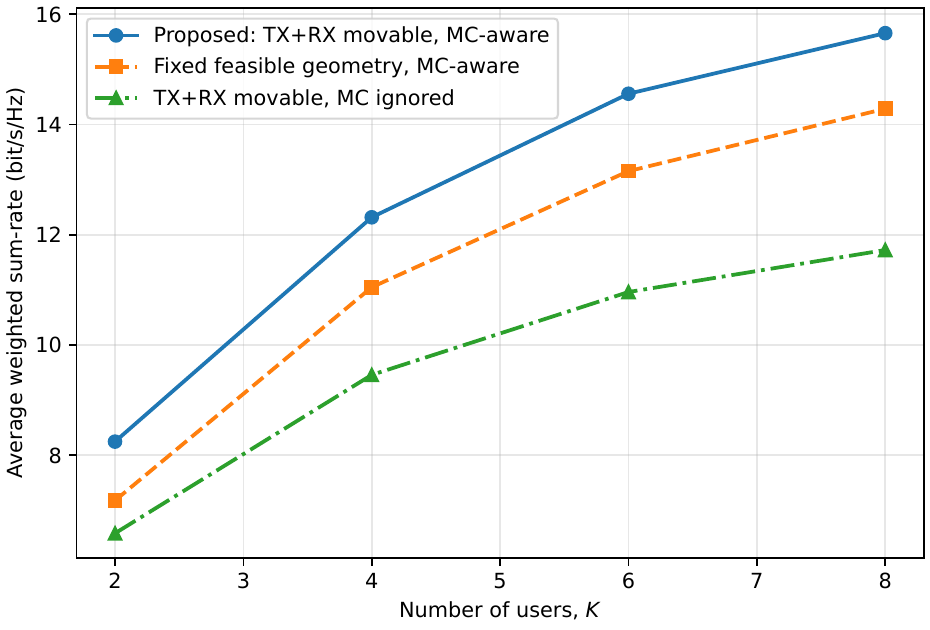}
 \caption{Average WSR versus the number of users at \SI{0}{dB} reference SNR
 with $M=16$.}
 \label{fig:user_scaling}
\end{figure}

\subsection{Ablation of Transmit- and Receive-Side Antenna Movement}
Table~\ref{tab:antenna_movement_ablation_grid} compares joint position optimization
with the fixed-array, TX-only, and RX-only designs for $M\in\{16,20,24\}$ and
$K\in\{4,8\}$ at \SI{0}{dB}. The joint design achieves the highest average WSR
in all six settings.

\begin{table}[t]
\centering
\caption{Average WSR (bit/s/Hz) for antenna-movement configurations at \SI{0}{dB} reference SNR.}
\label{tab:antenna_movement_ablation_grid}
\begin{tabular}{ccrrrr}
\toprule
$M$ & $K$ & Fixed & TX-only & RX-only & Joint \\
\midrule
16 & 4 & 10.997 & 11.070 & 12.044 & \textbf{12.215} \\
16 & 8 & 14.285 & 14.370 & 15.432 & \textbf{15.656} \\
\addlinespace
20 & 4 & 11.685 & 11.690 & 12.788 & \textbf{12.803} \\
20 & 8 & 15.289 & 15.295 & 16.504 & \textbf{16.521} \\
\addlinespace
24 & 4 & 11.925 & 11.928 & 13.045 & \textbf{13.056} \\
24 & 8 & 15.610 & 15.614 & 16.864 & \textbf{16.875} \\
\bottomrule
\end{tabular}
\end{table}

Among the single-block variants, RX-only position optimization captures most of the \textcolor{black}{gain from antenna movement}, whereas TX-only position optimization remains close to the fixed feasible grid. At
$M=16$, joint TX--RX position optimization still improves upon RX-only position optimization by $1.4\%$
for both user loads. The incremental gain decreases to $0.06\%$--$0.12\%$ for
$M\in\{20,24\}$, although the joint design remains the best in every setting.
Thus, receive-side movement provides most of the antenna-movement gain for the
larger compact transmit arrays considered here, while jointly optimizing both
position blocks yields the highest WSR in these settings.

\subsection{Impact of the Transmit and Receive Apertures}

Fig.~\ref{fig:aperture} separately varies the transmit- and
receive-aperture scales, with the other aperture held at its nominal size.
\textcolor{black}{The same feasible initial antenna positions are used across aperture scales.} The fixed-array
baseline remains at $8.251$ bit/s/Hz because its \textcolor{black}{antenna positions} do not change.

In Fig.~\ref{fig:aperture}\subref{fig:aperture_tx}, enlarging the transmit aperture scale from
$0.8$ to $2$ increases the WSR of the MC-aware joint design from $9.595$ to $9.982$ bit/s/Hz,
with most of the gain already realized by scale $1.5$. In contrast,
Fig.~\ref{fig:aperture}\subref{fig:aperture_rx} shows a steadier increase from $9.689$ to
$10.232$ bit/s/Hz as the receive aperture is enlarged. At scale $2$, the MC-aware joint design
outperforms the fixed grid by $21.0\%$ in the transmit-aperture sweep and by
$24.0\%$ in the receive-aperture sweep. The MC-ignored design benefits from a
larger feasible region but remains well below the MC-aware joint design throughout
both sweeps. These results indicate that the nominal compact apertures already
capture a large fraction of the available \textcolor{black}{gain from antenna movement}, while additional
receive-side space remains particularly useful for jointly improving the
desired channel and the coupling-dependent noise covariance.

\subsection{Scaling With the Number of Transmit Antennas}

Fig.~\ref{fig:tx_scaling} studies the effect of increasing $M$ under two
aperture policies. In Fig.~\ref{fig:tx_scaling}\subref{fig:tx_scaling_fixed_aperture}, the physical transmit
aperture is kept fixed. The WSR of the MC-aware joint design increases from $9.154$ bit/s/Hz at
$M=8$ to $13.056$ bit/s/Hz at $M=24$, but the incremental gain gradually
decreases as the aperture becomes crowded. The MC-ignored design saturates
much earlier: its WSR peaks around $M=16$ and remains near $9$ bit/s/Hz for
$M=20$ and $24$. Consequently, the gain of the MC-aware joint design over the MC-ignored design
widens from $10.4\%$ at $M=8$ to $44.9\%$ at $M=24$. \textcolor{black}{This widening gap highlights the increasing importance of accounting for coupling in the design when more antenna elements are packed into the fixed aperture.}

In Fig.~\ref{fig:tx_scaling}\subref{fig:tx_scaling_fixed_density}, both linear dimensions of the transmit
aperture are scaled by $\sqrt{M/8}$, so that the aperture area per transmit
antenna remains constant. All three designs then continue to improve with $M$.
At $M=24$, the proposed, fixed-grid, and MC-ignored designs achieve $14.712$,
$13.466$, and $12.132$ bit/s/Hz, respectively. Comparing the two panels shows
that saturation at large $M$ under a fixed aperture is consistent with the effects
of denser antenna packing and stronger MC. MC-aware position optimization
mitigates this effect, while enlarging the physical aperture provides an additional gain.

\subsection{Scaling With the Number of Users}

Fig.~\ref{fig:user_scaling} reports the average WSR for $K\in\{2,4,6,8\}$,
with $M=16$ and the nominal departure directions uniformly spread from
$-60^\circ$ to $60^\circ$. The aggregate WSR increases with $K$ for all designs. The MC-aware joint design remains above both baselines, and its advantage over the MC-ignored movable design grows from $25.3\%$ at $K=2$ to $33.6\%$ at $K=8$.

\section{Conclusion}\label{sec:conclusion}
\textcolor{black}{This paper investigated joint precoding and antenna positioning for MA-enabled MU-MIMO systems with position-dependent MC. We combined field-response and circuit-based models and developed a current-domain WMMSE framework with MM-based position updates and a conic ADMM solver. Numerical results demonstrated the WSR benefits of the proposed design over both MC-aware fixed arrays and movable arrays designed without considering MC. The current-domain formulation also clarifies the different roles of transmit and receive antenna positioning: transmit positions affect the power required to generate the antenna currents, while receive positions affect the noise covariance, alongside their respective effects on propagation. The impact of these coupling-dependent quantities is governed by the transmit-current covariance and receive processing. Consequently, the antenna positions should be optimized together with the precoders and receive processing, accounting for their combined effects on desired signals, interference, transmit power, and receive noise.}

\FloatBarrier

\appendix
\subsection{Proof of Proposition \ref{prop:phase_majorizer}}
\label{append:phase_majorizer_proof}
For any $\mathbf p_{\mathtt T}\in\mathcal A_{\mathtt T}^{M}$, the bounds in \eqref{eq:aperture_phase_extrema} place the phase $\mathbf a_q^{\transpose}\mathbf p_{\mathtt T}+\arg(\chi_q)$ in $\overline\Theta_q$. The definition of $\overline\mu_q$ therefore bounds the scalar coefficient in \eqref{eq:harmonic_exact_hessian_appendix} from above, and, since $\mathbf a_q\mathbf a_q^{\transpose}\succeq\mathbf 0$, we obtain $\nabla^2 h_q(\mathbf p_{\mathtt T})\preceq\overline\mu_q\mathbf a_q\mathbf a_q^{\transpose}$. Summing these inequalities over $q\in\mathcal H$ gives the Hessian bound in \eqref{eq:phase_hessian_upper_bound}. Moreover, convexity of $\mathcal A_{\mathtt T}$ ensures that the line segment from $\overline{\mathbf p}_{\mathtt T}$ to any feasible $\overline{\mathbf p}_{\mathtt T}+\boldsymbol\delta_{\mathtt T}$ lies entirely in $\mathcal A_{\mathtt T}^{M}$, so the same Hessian bound holds throughout the segment. Applying this bound to the integral remainder in Taylor's theorem bounds the remainder by $\frac12\boldsymbol\delta_{\mathtt T}^{\transpose}\overline{\mathbf M}_{\mathrm{ph}}\boldsymbol\delta_{\mathtt T}$, where the factor $\frac12$ follows from $\int_0^1(1-t)\,\mathrm dt=\frac12$. Substituting this bound into the Taylor expansion gives \eqref{eq:phase_majorization_inequality}. This completes the proof.

\subsection{Proof of Proposition \ref{prop:coupling_majorizer}}
\label{append:coupling_majorizer_proof}
Because \(\mathbf C_{\mathtt T}\) is real symmetric and \(\overline{\mathbf S}\) is Hermitian, the accepted transmit power can be written as
\begin{equation}\label{eq:coupling_graph_edge_sum}
	\mathcal P_{\mathtt T}(\mathbf p_{\mathtt T})
	=\mathrm{constant}+2\sum_{m<n}a_{\mathtt T,mn}c_{\mathtt T}(r_{mn}).
\end{equation}
Differentiating the edge potentials gives the gradient in
\eqref{eq:coupling_graph_gradient}. For one edge, the Hessian with respect to
\(\mathbf d_{mn}=\mathbf p_{\mathtt T,m}-\mathbf p_{\mathtt T,n}\), with
\(r=\|\mathbf d_{mn}\|_2\) and \(\mathbf u=\mathbf d_{mn}/r\), is
\begin{equation}\label{eq:coupling_edge_hessian_appendix}
	\mathbf{W}_{mn}(r)
	=2a_{\mathtt T,mn}\left[
	c_{\mathtt T}''(r)\mathbf{u}\mathbf{u}^{\transpose}
	+\frac{c_{\mathtt T}'(r)}{r}
	(\mathbf{I}_2-\mathbf{u}\mathbf{u}^{\transpose})
	\right].
\end{equation}
Its radial and tangential eigenvalues are
\(2a_{\mathtt T,mn}c_{\mathtt T}''(r)\) and
\(2a_{\mathtt T,mn}c_{\mathtt T}'(r)/r\), respectively, so
\eqref{eq:adaptive_edge_curvature} gives
\begin{equation*}
	\mathbf{W}_{mn}(r)\preceq
	\overline{\omega}_{mn}\mathbf{I}_2,
	\qquad r\in\overline{\mathcal{R}}_{mn}.
\end{equation*}
The contribution of this edge to the stacked position Hessian is
\begin{equation*}
	(\mathbf{e}_m-\mathbf{e}_n)
	(\mathbf{e}_m-\mathbf{e}_n)^{\transpose}
	\otimes\mathbf{W}_{mn}(r).
\end{equation*}
Because \((\mathbf e_m-\mathbf e_n)(\mathbf e_m-\mathbf e_n)^{\transpose}\succeq\mathbf0\), summing the edge bounds gives \(\nabla^2\mathcal P_{\mathtt T}\preceq\overline{\mathbf M}_{\mathtt T}^{\mathcal P}\) whenever all pairwise distances lie in their respective intervals \(\overline{\mathcal R}_{mn}\). The distance condition in Proposition~\ref{prop:coupling_majorizer} ensures that this Hessian bound holds throughout the update segment, so Taylor's theorem with integral remainder gives \eqref{eq:coupling_majorization_inequality}. This completes the proof.

\bibliographystyle{IEEEtran}
\bibliography{reference}
	
\end{document}